\documentclass{aa}  

\usepackage{graphicx}
\usepackage{txfonts}
\usepackage[]{hyperref}
\begin{document}

   \title{
     Improving 1D stellar atmosphere models with insights from multi-dimensional simulations  }

   \subtitle{I. 1D vs 2D stratifications and spectral comparison for O stars}

\author{G.\ Gonz\'alez-Tor\`a\inst{\ref{inst:ARI}}
        \and
        A.\ A.\ C.\ Sander\inst{\ref{inst:ARI}}
        \and 
        J.\ O.\ Sundqvist\inst{\ref{inst:KUL}}
        \and
        D.\ Debnath\inst{\ref{inst:KUL}}
        \and 
        L.\ Delbroek\inst{\ref{inst:KUL}}
        \and 
        J.\ Josiek\inst{\ref{inst:ARI}}
        \and 
        R.\ R.\ Lefever\inst{\ref{inst:ARI}}
        \and
        N.\ Moens\inst{\ref{inst:KUL}}
        \and 
        C.\ Van der Sijpt\inst{\ref{inst:KUL}}
        \and
        O.\ Verhamme\inst{\ref{inst:KUL}}
        }

   \institute{{Zentrum für Astronomie der Universität Heidelberg, Astronomisches Rechen-Institut, 
   Mönchhofstr. 12-14, 69120 Heidelberg\label{inst:ARI}}\\
        \email{gemma.gonzalez-tora@uni-heidelberg.de}
   \and
        {Institute of Astronomy, KU Leuven, Celestijnenlaan 200D, 3001, Leuven, Belgium\label{inst:KUL}}\\
}  

   \date{Received \today; accepted -}

% \abstract{}{}{}{}{} 
% 5 {} token are mandatory
 
  \abstract
   {The outer layers and the spectral appearance of massive stars are inherently affected by radiation pressure. Recent multi-dimensional, radiation-hydrodynamical (RHD) simulations of massive stellar atmospheres have shed new lights on the complexity involved in the surface layers and the onset of radiation-driven winds. These findings include the presence of subsurface %subsonic, 
   radiatively driven turbulent motion. For some regimes, the velocities associated with this turbulence and their localization significantly exceed earlier estimates from stellar structure models, prompting the question whether spectral diagnostics obtained with the typical assumptions in 1D, spherically symmetric and stationary atmospheres are still sufficient. 
   }
   { 
   For the foreseeable future, the inherent computation costs and necessary approximations will make the common usage of multi-dimensional, time-dependent atmosphere models in quantitative spectral analysis of populations of stars very difficult. Therefore, suitable approximations of multi-dimensional simulation results need to be implemented into 1D atmosphere models.
   }
   {We compare current 1D and multi-dimensional atmosphere modelling approaches to understand their strengths and shortcomings. We calculate averaged stratifications from selected multi-dimensional calculations for O stars -- corresponding to the spectral types O8, O4, and O2, with $\log g\sim3.7$ -- to approximate them with 1D stellar atmosphere models using the PoWR model atmosphere code and assuming a fixed $\beta-$law for the wind regime. We then study the effects of our approximations and assumptions on current spectral diagnostics. In particular, we focus on the impact of an additional turbulent pressure in the subsonic layers of the 1D models.
   }
   {To match the 2D averages, the 1D stellar atmosphere models need to account for turbulent pressure in the hydrostatic equation. Moreover, an adjustment of the connection point between the (quasi-)hydrostatic regime and the wind regime is required. The improvement between the density stratification of 1D model and 2D average can be further increased if the mass-loss rate of the 1D model is not identical to those of the 2D simulation, but typically $\sim0.2\,$dex higher. Especially for the early type star, this implies a significantly more extended envelope with a lower effective temperature.}
   {Already the inclusion of a constant turbulence term in the solution of the hydrostatic equation sufficiently reproduces the 2D-averaged model density stratifications. The addition of a significant turbulent motion also smoothens the slope of the radiative acceleration term in the (quasi-)hydrostatic domain, with several potential implications on the total mass-loss rate inferred from 1D modelling. Concerning the spectral synthesis, the addition of a turbulence term in the hydrostatic equation mimics the effect of a lower surface gravity, potentially presenting a solution to the ``mass discrepancy problem'' between the evolutionary and spectroscopy mass determinations.   
   }

   
   \maketitle

\section{Introduction}\label{sec:intro}
Massive stars ($M\gtrsim8\,M_{\odot}$) display strong stellar winds that are critical to understand their evolutionary path and impact, ranging from their immediate fate up to the chemical enrichment of galaxies \citep[e.g.,][]{Portinari+1998}. The strong winds inject momentum into the surrounding interstellar medium (ISM) and enrich the ISM with nuclear processed material. Hot, massive stars show a UV-dominated radiation field, which is not only capable of ionizing the ISM, but also provides the decisive source of radiation pressure that is able to drive their stellar winds by momentum transfer from the photons on metal ions by absorption and scattering in line transitions \citep[e.g.,][]{Castor+1975,Pauldrach+1986}.

To derive the properties of hot massive stars, stellar atmosphere models are crucial. In particular for the hottest stars, the information about the temperature and surface gravity is only encoded in the spectral lines as other diagnostics such as the Balmer jump are no longer visible. Moreover, the imprint of the stellar winds onto the resulting spectral lines has to be taken into account. Given the inherent deviation from Local Thermodynamical Equilibrium (LTE) and the need for a sophisticated radiative transfer to account for the line-specific interactions, the standard framework for the computation of hot star atmospheres is already complex and numerically challenging \citep[see e.g.,][]{Hillier2003,HubenyLanz2003,HamannGraefener2003}. Consequently, detailed atmosphere modelling for the spectral synthesis of O, B, and Wolf-Rayet (WR) stars is essentially performed exclusively in 1D  \citep{Puls2008,Sander2017}. With this approach, state-of-the-art expanding atmosphere codes such as CMFGEN \citep{HillierMiller1998}, PoWR \citep{Graefener+2002} and FASTWIND \citep{Puls+2005} are widely applied to perform quantitative spectral analysis in various environments \citep[e.g.,][]{Hillier+2003,Najarro+2011,Bouret+2012,RamirezAgudelo+2017,Ramachandran+2018,Carneiro+2019,Hainich+2020} with different metallicities ($Z)$. In the limit of negligible wind impact on the spectrum, also the static code TLUSTY is applied \citep[e.g.,][]{Bouret+2003,Marcolino+2009}.

Despite the success of 1D stellar atmosphere models, there are also clear indications that the atmospheres and winds of hot stars do not fully adhere to the idealized assumptions of spherical symmetry with a stationary outflow \citep[e.g.,][]{OwockiPuls1999,DessartOwocki2005,Sundqvist+2010,Surlan+2012,Sundqvist+2018,Hennicker+2018}. Phenomena such as convection or wind clumping are inherently three-dimensional. Current 1D treatments for various processes are often ad-hoc descriptions with limited physical insights on the underlying process \citep[see e.g.,][for the case of wind clumping]{Hawcroft+2021,Ruebke+2023,Verhamme+2024}. On the other hand, a full 3D treatment of a non-LTE, time-dependent, expanding stellar atmosphere is so far computationally unfeasible. However, new insights can emerge from exploratory 2D and 3D models, which will then have to be sufficiently approximated and parameterized in 1D atmosphere calculations.

In this work, we focus on the regime of O (super-)giants and in particular the treatment of their sub-surface structure. In this regime, sub-surface turbulence 
is predicted to arise due to the \textit{iron-opacity peak} leading to a very unstable sub-surface layer \citep{Jiang+2015}, characterised by large turbulent velocities and significant effects on spectral absorption line diagnostics \citep{Schultz+2022}.  
Based on a newly established suite for time-dependent radiation-hydrodynamic (RHD) simulations \citep{Moens+2022fld} and a hybrid-opacity approach developed by \citet{Poniatowski+2022}, first applied in the field of WR stars \citep{Moens+2022wr}, \citet{Debnath+2024} established pioneering 2D unified atmosphere and wind models covering the outer stellar layers from below the iron-opacity peak region up to the inner layers of the radiation-driven outflow. They found significant sub-photospheric dynamics, which -- to first degree -- they suggested to be parameterized as a turbulent velocity.

The inclusion of turbulence in the hydrostatic solution of a 1D hot star atmosphere is currently not common in the study of O stars \citep[though see, e.g.,][for B supergiants]{Wessmayer+2022}. However, the addition of an ad-hoc photospheric 'microturbulence' and 'macroturbulence' as an extra-line broadening mechanism is common to fit the observations of photospheric absorption lines of O stars \citep{Howarth+1997, Simon-Diaz+2017}. Such studies find a photospheric macroturbulent velocity in a good agreement with the turbulent velocities found by \citet{Debnath+2024}. In the framework of the X-Shooting ULLYSES \citep[XShootU,][]{Vink+2023} collaboration, \citet{Sander+2024} recently analyzed three O stars comparing different methods using CMFGEN, PoWR, and FASTWIND. While yielding an overall good agreement for the derived parameters, one of the outcomes of their study is that the inclusion of turbulence in the calculation of the hydrostatic structure requires a deeper investigation and might even be a promising tool to address the long-standing issue of the so-called ``mass discrepancy'' \citep{Herrero+1992}, denoting a notable difference between the masses derived from quantitative spectroscopy with those predicted by stellar evolution models.

To study the implications from the results by \citet{Debnath+2024}, we calculate 1D stellar atmosphere models with the PoWR model atmosphere codes, which allows for the explicit inclusion of a turbulent velocity in the hydrostatic solution, and compare them to the spatially averaged 2D model profiles of three O-type (super-)giant stars from \citet{Debnath+2024}. We carefully study the effect of varying several input parameters and show how the derived stellar properties are affected when accounting for the structural predictions from the 2D models. This is the first study that aims to compare more systematically the short-comings and strengths of both methodologies.                                                                    

\medskip
This paper is organized as follows: in Section~\ref{sec:methods} we present the main characteristics of the 1D framework used, PoWR, as well as highlight the main differences with the multi-dimensional framework. 
Section~\ref{sec:results} shows the results for the different profile comparisons as well as the spectral synthesis. We discuss the implications of our results in Section~\ref{sec:diss}. Finally we conclude our work in Section~\ref{sec:conc}.

%--------------------------------------------------------------------

 \section{Methods}\label{sec:methods}

\subsection{The PoWR 1D stellar atmosphere code}
The Potsdam Wolf-Rayet stellar atmosphere code \citep[PoWR,][]{Graefener+2002,HamannGraefener2003,Sander+2015} calculates stationary, non-LTE atmospheric models assuming spherical symmetry. The population numbers are calculated from the assumption of statistical equilibrium \citep[e.g.,][]{Hamann1986} and the radiative transfer is solved in a unified co-moving frame (CMF) scheme for continuum and lines which makes use of a two-step approach with a ray-by-ray and a momentum-solution of the radiative transfer equation \citep{Koesterke+2002}. The temperature stratification is solved using either a generalized form of the Uns{\"o}ld-Lucy approach \citep{HamannGraefener2003} or the thermal balance of electrons \citep{Kubat+1999,Sander2015}. The multitude of iron levels and transitions is accounted for with a super-level approach combining the levels of the iron group elements (Sc to Ni) into energy bands \citep{Graefener+2002} and nowadays also different parity.

In PoWR, the radius $R_\ast$ describes the inner boundary of the model where the specified maximum Rosseland continuum optical depth is reached, which we typically set to $\tau_{\mathrm{Ross,cont}}=20$. Motivated by its roots, PoWR does define an effective temperature $T_\ast$ corresponding to $R_\ast$, which can differ from the more common definition of $T_{\mathrm{eff}}$ as the effective temperature at $\tau_{\mathrm{Ross}}=2/3$. While the differences for the models in this work are not large due to their moderate mass-loss rates, we refrain from listing $T_\ast$ and only report the common $T_{\mathrm{eff}} \equiv T_{2/3} := T_{\mathrm{eff}}(\tau_{\mathrm{Ross}}=2/3)$ to enable an easier comparison.

An optional branch of the code allows to couple the wind parameters to the stellar parameters by solving the hydrodynamic equation of motion \citep{Sander+2017,Sander+2023}, but the default is the assumption of an analytic velocity law
\begin{equation}
  \label{eq:vbeta}
    \varv(r)=p_{1}\left( 1-\frac{1}{r+p_{2}} \right)^\beta
\end{equation}
with the parameters $p_{1}$ and $p_{2}$ being fixed by two boundary conditions, namely $\varv(r_{\mathrm{max}})=\varv_{\infty}$ and $\varv(r_{\mathrm{con}})=\varv_{\mathrm{con}}$. Here, $r_{\mathrm{con}}$ is the connection point between the hydrostatic and the wind regime. Different settings are possible for this connection point. In this work, we set $r_{\mathrm{con}}$ to the radius where the velocity reaches a factor $0.95$ of the local effective sound speed accounting for the turbulence. The implications on changing the connection point are discussed in Section\,\ref{sub:vel}. The terminal velocity $\varv_{\infty}$ and $\beta$ are free parameters to be provided by the user. 
For O stars, $\beta=0.8$ is initially assumed, as motivated by the later extensions \citep{Pauldrach+1986} of the CAK theory \citep*[named after][]{Castor+1975}. However, the assumption of a $\beta$-law has several consequences not only on the velocity profile but also the emergent spectra \citep[e.g.,][]{Lefever+2023}. In this work, we vary $\beta$ to better match 2D results (see Section~\ref{sec:results}). 

From $\varv(r)$, the density stratification $\rho(r)$ is established via a fixed (input) mass-loss rate $\dot{M}$ and the equation of continuity
\begin{equation}
  \label{eq:conteq}
   \dot{M} = 4 \pi r^2 \varv(r) \rho(r)\text{.}
\end{equation}
In the subsonic, (quasi-)hydrostatic regime below $r_{\mathrm{con}}$, the density and velocity are obtained by the direct numerical integration of the hydrostatic equation
\begin{equation}
   \label{eq:hydrostat}
    \frac{\mathrm{d}P}{\mathrm{d}r}=-\rho(r) \left[ g(r)-a_{\mathrm{rad}}(r) \right]
\end{equation}
including the total radiative acceleration $a_{\mathrm{rad}}(r)$ calculated in the CMF \citep{Sander+2015}, which yields the velocity $\varv_{\mathrm{con}}$ where the connection to the $\beta$-law regime is established.  

To connect density and pressure, we require an equation of state. For this we assume the usual ideal gas equation $P(r)=\rho(r)a_\mathrm{s}^{2}(r)$, however with $a_\text{s}$ being an effective speed that includes both thermal sound speed and turbulence $\varv_\mathrm{turb}$, i.e.,
\begin{equation}
    \label{eq:as}
    a_\text{s}^{2}(r)=\frac{k_{\mathrm{B}}T(r)}{\mu(r)m_{\mathrm{H}}}+ \varv_{\mathrm{turb}}^{2}(r)\text{.}
\end{equation}
In Eq.\,\eqref{eq:as}, $T(r)$ denotes the electron temperature and $\mu(r)$ the mean particle mass while the constants $k_\mathrm{B}$ and $m_\mathrm{H}$ have their usual meaning as Boltzmann's constant and the atomic hydrogen mass. Any given non-zero value for the \textit{turbulent velocity},  $\varv_{\mathrm{turb}}$, forces the inclusion of a turbulent pressure $P_\mathrm{turb}(r) = \rho(r) \varv_{\mathrm{turb}}^{2}(r)$ in the solution of the hydrostatic equation, of which we will make extensive use in this work. 

The so-far introduced $\varv_{\mathrm{turb}}$ is a value that formally only affects the solution of the hydrostatic equation. The line opacity profiles in the radiative transfer calculations needed to compute the atmospheric structure are instead calculated with a Doppler profile assuming a depth-independent $\varv_\mathrm{dop}$ value which we set to $30\,\mathrm{km}\,\mathrm{s}^{-1}$ for all models presented in this work. This is a typical value used in the PoWR OB-star grids \citep{Hainich+2019}.

Finally, the emergent spectrum is calculated with a final radiative transfer calculation in the observer's frame. For these, the line opacities are modelled with a detailed, depth- and element-dependent Doppler profile accounting for both thermal and microturbulent broadening \citep{Shenar+2015}. We denote the corresponding microturbulent velocity for this observer's frame calculation as $\xi(r)$ throughout this work. While $\xi(r)$ is expected to increase in the wind, its minimum (i.e., ``photospheric'') value $\xi_\mathrm{min}$ can, but does not have to, be set to the same value as the $\varv_{\mathrm{turb}}$ included in the hydrostatic equation. In our effort to approximate sub-surface motions via a turbulent pressure, we will make explicit use of different values for $\xi_\mathrm{min}$ and $\varv_{\mathrm{turb}}$.

To account for wind inhomogenieties, PoWR can use the so-called \textit{microclumping} approach assuming small scale, optically thin clumps that follow the smooth atmospheric velocity field surrounded by a void interclump medium. In this work, we do not focus on any clumping impact. In this setting, motivated by the findings in multi-dimensional simulations that question the standard two-component medium approach \citep{Moens+2022wr,Debnath+2024}, we have assumed smooth models with density contrast $D = 1$.  
Despite not performing a detailed study on the influence of the clumping factor, we compare the smooth model profile with a $D = 10$ model profile in Appendix~\ref{app:clump}, where no clumping is assumed until the sonic point with a cosine-shaped transition to a maximum density contrast $D = 10$, which is kept constant beyond $10\,R_\ast$. There is no significant difference between both models. 

We further assumed the solar abundances from \citet{Asplund+2009} for all our 1D models, while \citet{Debnath+2024} used the set from \citet{GrevesseNoels1993}. We do not expect any significant impact of this abundance difference for any of our comparisons. 
The impact on the radiative force in the hydrostatic domain is expected to be lower than the influence of other choices such as the assumption $\varv_\mathrm{dop}$. For the spectral synthesis, we only compare spectra from PoWR models using the same abundance sets.

 \subsection{2D radiation-hydrodynamic framework}\label{sec:2d}

In this section, we focus on the main differences between the 1D PoWR models and the multi-dimensional radiation-hydrodynamic (RHD) models from \citet{Debnath+2024}. A more extensive description of the multi-dimensional atmospheric framework can be found in \citet{Moens+2022wr,Moens+2022fld} as well as \citet{Debnath+2024}.

The multi-dimensional approach is based on the MPI-AMRVAC code  \citep{Xia+2018} which solves the partial differential equations (PDEs) on a finite volume grid. We use the RHD module from \citet{Moens+2022fld} in a `box-in-a-star' approach including corrections to account for the spherical divergence \citep{Sundqvist+2018, Moens+2022fld}. Opacities are calculated using the method by \citet{Poniatowski+2022}. We compare with the O-star models from \citet{Debnath+2024}, where the finite volume extends from the lower boundary inside the stellar envelope ($R_{o}$, located at $T_{0} \approx 460$, 470 and 480 kK for the O8, O4 and O2 models, respectively).
This comprises the optically thick inner region of the stellar atmosphere, the launching region of the wind, and the supersonic outflow.  

The 1D PoWR code solves the frequency-dependent radiative transfer equation in the CMF. The velocity entering the 1D radiative transfer equations is the previously described connection of the solution of the hydrostatic equation and a $\beta-$law. A closure for the solution of moment equations is provided by two Eddington factors ($f_\nu$, $g_\nu$) relating different moments of the radiation field with initial values for $f_\nu$ and $g_\nu$ being obtained from a ray-by-ray radiative transfer calculation \citep{Koesterke+2002}. For the multi-dimensional approach, however, the Eddington factors become tensors and their direct calculation would require a frequency- and angle-dependent radiative transfer calculation in three dimensions, which even nowadays would be computationally prohibitive. 
Instead, the multi-dimensional framework solves only the frequency-integrated 0th moment equation of the time-dependent radiative transfer equation -- describing the conservation of radiation energy density -- and applies a closure relation based on a flux-limiting diffusion (FLD) approximation. This approach recovers the correct limits for radiative energy density and pressure in the optically thick and optically thin regimes, 
but uses an approximate analytic law to bridge these two limits \citep{Pomraning+1988}, simplifying the radiation moment equations.  
The reader is referred to \citet{Moens+2022fld} for an extensive description of the FLD method.

The studied multi-dimensional models include deeper layers where the opacity increases due to the iron recombination (\textit{iron-opacity peak} region), while the 1D PoWR models presented in this work do not go as deep ($T^{1D}_{0}\approx70$, 80 and 90 kK for the O8, O4 and O2 models, respectively). As mentioned above, the 2D models find large turbulent motions in layers well beneath the photosphere caused by the radiative force arising from the increased opacity. Depending on the specific stellar parameters, \citet{Debnath+2024} derived turbulent velocities of $\sim 30-100$ km\,s$^{-1}$.

The deep origin of the turbulence raises the question whether it might no longer be sufficient to follow the standard 1D atmosphere model approach which usually limits the treatment of the subsurface layers to maximum optical depths of $\sim$$10 \dots 100$. In the bulk of PoWR models, the lower boundary is set to $\tau_{\mathrm{Ross,cont}} = 20$ \citep[though see, e.g.,][for higher and lower examples]{SanderVink2020,Sabhahit+2023}. This is considered more than sufficient for the aim to derive synthetic spectra, but this only holds if the presumed density stratification in the (quasi-)hydrostatic regime is correct. While we aim to eventually explore deeper 1D models covering the iron-opacity peak as they have been calculated for WR stars \citep[e.g.,][]{Sander+2020,Sander+2023}, we will explore in this work whether the standard treatment can be kept by simply including the contribution of a turbulent velocity in the hydrostatic equation (Eq.~\ref{eq:hydrostat}).

We note that in our 1D PoWR approach we neglect potential effects from energy transport by enthalpy (`convection') in our 1D models. As shown in \citet{Debnath+2024}, however, for the O-star models studied here such convective energy transport reaches maximum $\sim$$10\,\%$ of the total luminosity at the peak of the iron-bump and decreases rapidly when approaching the overlying photosphere. As such, our assumption of a purely radiative \textit{energy} transport for the 1D PoWR models is well justified in this regime.

In the multi-dimensional RHD models, a hybrid opacity approach is used \citep{Poniatowski+2022} to accelerate the gas from the hydrostatic regime (`core'). In this formalism, opacities are estimated using tabulated Rosseland mean opacities from OPAL \citep{IglesiasRogers1996} combined with enhanced line opacities described by self-consistently calculated CAK-like force multipliers using the concepts introduced by \citet{Gayley1995}. This treatment is motivated by the fact that static Rosseland mean opacities are sufficient to lift up the gas from the hydrostatic stellar core, while the enhanced line driving opacities due to Doppler shifts have to be taken into account once the gas is further away in an optically thin supersonic regime. 

 In the stationary 1D atmosphere models, the opacities are not tabulated, but directly calculated from the non-LTE population numbers \citep[with a superlevel approach for iron and iron group elements (Sc to Ni), see][]{Graefener+2002}. With the calculation of the radiative transfer in the CMF, the proper 1D radiative acceleration is directly obtained for the assumed velocity field with no further simplifications. Once the iteration for a stationary solution between the radiation field from the CMF as well as the non-LTE level populations is obtained, the converged values are used to compute the spectrum in the 1D models.%

In contrast, the multi-D models assume LTE-level populations to compute force multipliers and flux-weighted frequency-integrated opacities in the outer wind. Furthermore, for practical reasons, both the energy- and Planck-mean opacities are assumed to be equal to this flux mean, which very likely significantly overestimates heating and cooling effects in the stellar wind. This renders a wind structure that is practically in radiative equilibrium  
\citep[see discussion in Section 5.5 of][]{Debnath+2024}. 
Further efforts that aim at a description of non-LTE effects as well as a split treatment of the different opacity means in the multi-D models are currently being explored. 

Finally, a significant difference on the modelling approaches relates to the treatment of atmospheric inhomogenities. 
While the 1D models need to rely on ad-hoc assumptions such as photospheric macroturbulence or clumping, the multi-dimensional models are free from these prescriptions. 
 To describe wind inhomogeneities, 1D models usually implement a two-component medium (i.e., optically thin clumps surrounded by void), where only one of the components (i.e., the clumps) is explicitly described. In contrast, multi-dimensional studies \citep[e.g.,][]{Moens+2022fld,Moens+2022wr,Debnath+2024} have found that the wind-density is distributed around a mean, seriously questioning the standard 
 two-component medium approach. A more detailed exploration on clumping assumptions will be left to a future study, while we focus on the turbulent motion in this work.

 \section{Calculation and results}\label{sec:results}

\citet{Debnath+2024} use the multi-dimensional approach discussed in Section~\ref{sec:2d} to %reproduce the stellar 
compute atmospheric properties of three O-type supergiants. They obtain probability density maps and lateral-temporal unweighted spatial averages 
for the radial velocity, gas density and radiation temperature for O8, O4, and O2 models \citep[see Figure 10 in][]{Debnath+2024}. The fundamental parameters for the averaged 2D star models are also shown in Table 1 of \citet{Debnath+2024}.

We have calculated the PoWR stellar models to reproduce the unweighted 2D-averaged stellar parameter profiles such as the wind velocity, the density, and the gas temperature from the models in \citet{Debnath+2024}. The fitting was done by visually inspecting the averaged 2D profiles with the 1D profiles. The 2D models were designed to describe typical O8, O4 and O2-type (super-)giants.
Initially, we adopted the same luminosity ($\log(L_{\star}/L_{\odot})$), mass-loss rate ($\log\dot{M}/M_{\odot}\mathrm{yr}$), stellar mass ($M_{\star}/M_{\odot}$), and effective temperature ($T_{\mathrm{eff}}$) as the 2D-averaged fundamental parameters listed in Table 1 of \citet{Debnath+2024}. For the wind velocity law, we initially assume $\beta=0.8$.

As we describe in more detail below, the initial values are subsequently altered in order to better reproduce the 2D averages for the wind velocity, density, and gas temperature profiles. In Fig.\,\ref{fig:profile-O4}, we illustrate the process focusing on the O4 star case: The solid-black lines are the lateral- and temporal-averaged 2D model profiles from \citet{Debnath+2024}, while the blue lines mark the corresponding profiles of the 1D PoWR models using the the above mentioned initial set with (solid blue, $\varv^{\text{1D}}_{\mathrm{turb}}>0$) and without (dashed blue, $\varv^{\text{1D}}_{\mathrm{turb}}=0$) adding turbulent pressure to the hydrostatic equation. We then allow for variations in the mass-loss rate $\dot{M}$, $\beta$, and the connection settings to obtain an even better reproduction. The resulting ``best fit'' model is shown as a solid red line in Fig.~\ref{fig:profile-O4} with our derived $\varv_{\mathrm{turb}}>0$ and best-fit parameters shown in Table~\ref{tab:params}.

\subsection{Atmospheric structure profile comparison}\label{sec:comp}

    \begin{figure}
      \centering
      \includegraphics[width=1.\linewidth]{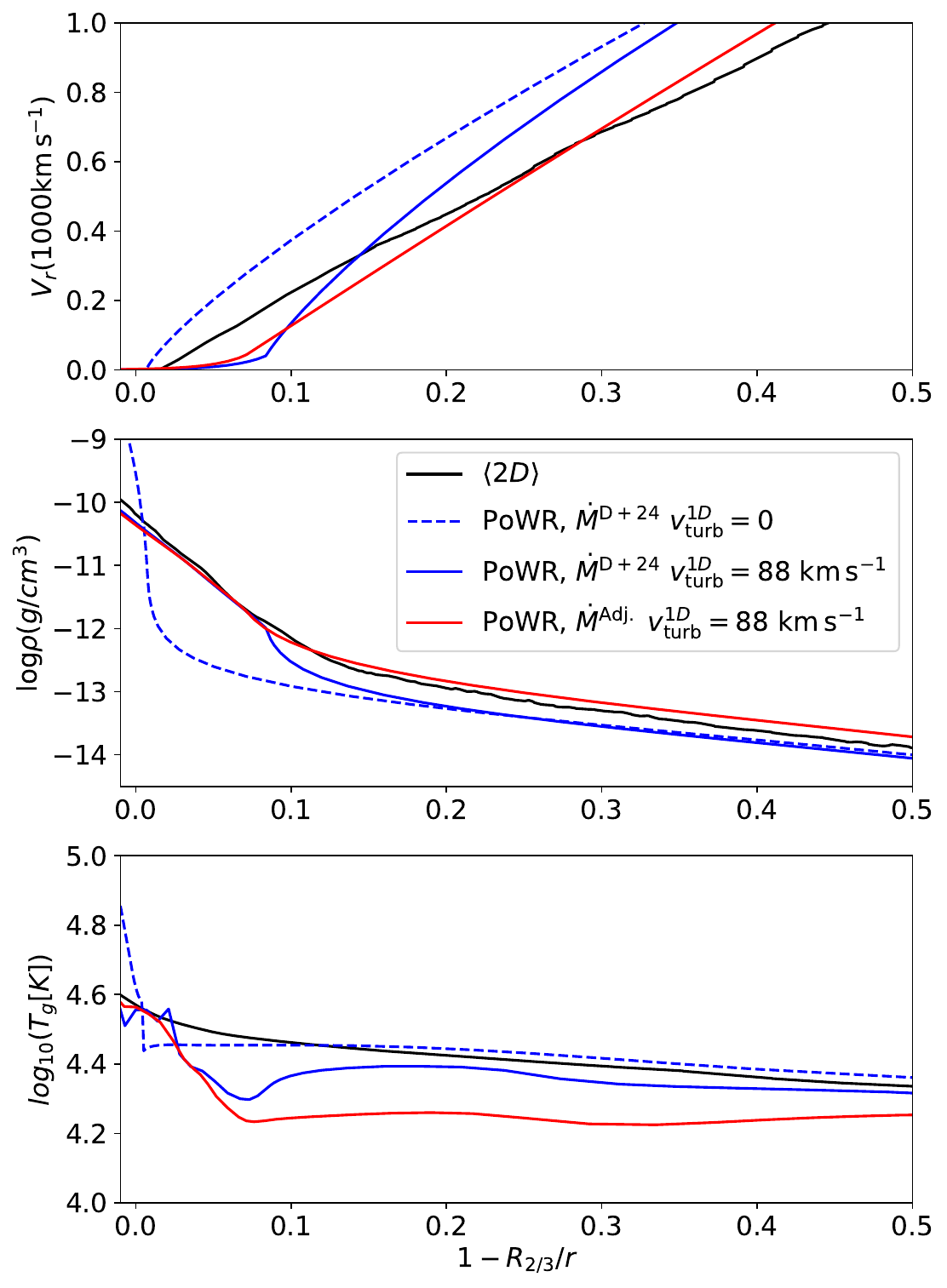}
      \caption{ Profile comparison for the O4 star. For visualization purposes, the 2D averaged model and 1D have been calibrated such that they have the same $R_{2/3}$. \textit{Upper panels: }Wind velocity profile, in solid black for the 2D spatially averaged model of \citet{Debnath+2024}, dashed blue for 1D PoWR model with $\varv^\text{1D}_{\mathrm{turb}}=0$ and solid blue for $\varv^\text{1D}_{\mathrm{turb}}=88$ km$\,$s$^{-1}$ with the same parameters as in Table 1 of \citet{Debnath+2024}, solid red for the 1D PoWR model with the final best-fit parameters from Table~\ref{tab:params} (i.e., adjusted $\dot{M}$ and $\beta-$law) and $\varv_{\mathrm{turb}}=88$ km$\,$s$^{-1}$. \textit{Middle panels: }Same as the \textit{upper panel} but for the density profile. \textit{Lower panel:} Same as the \textit{upper panel} but for the gas temperature. 
      }
      \label{fig:profile-O4}
  \end{figure}

\begin{figure*}
      \centering
      \includegraphics[width=1.\linewidth]{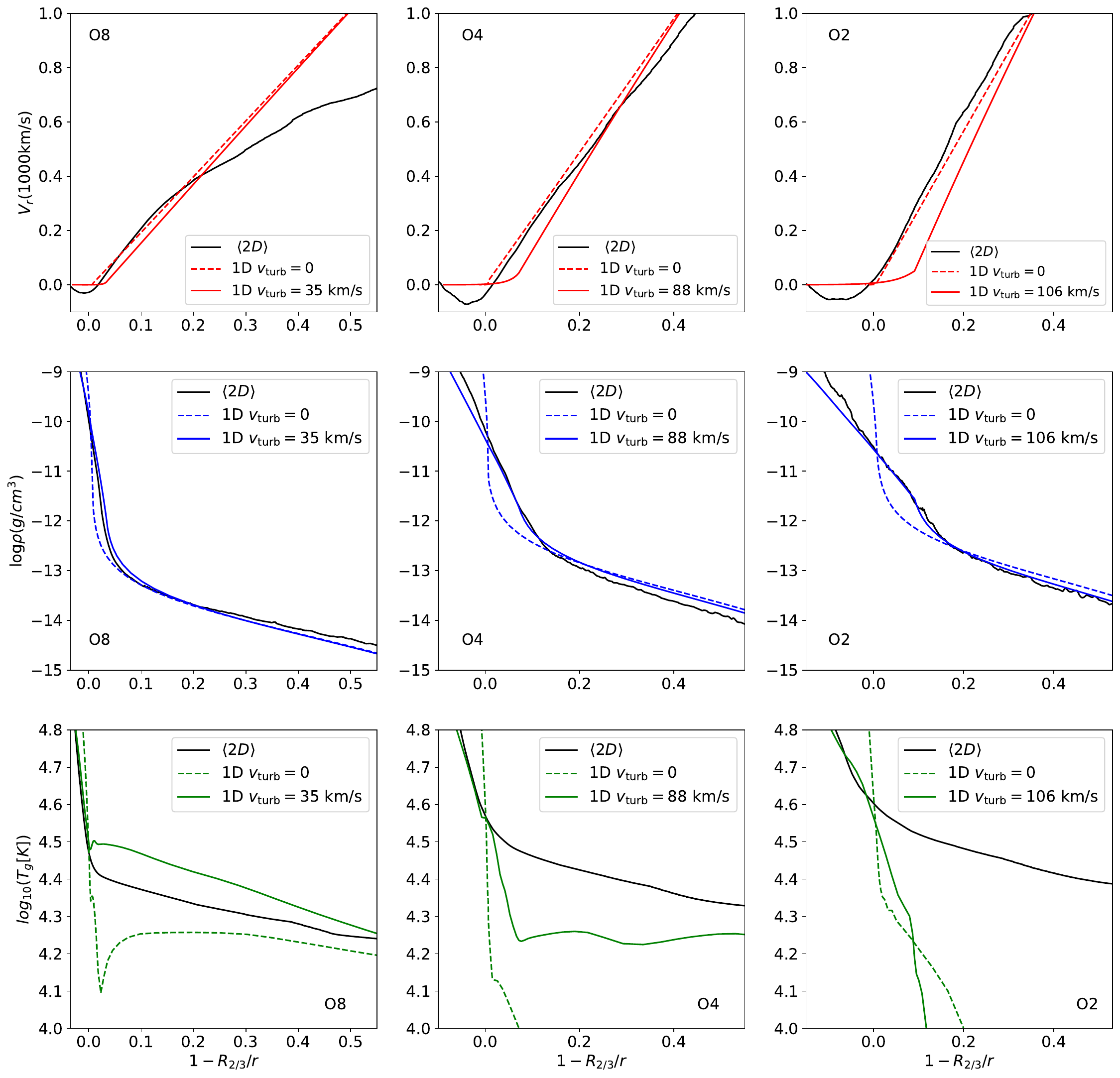}
      \caption{ Profile comparison for an O8, O4 and O2 stars. \textit{Upper panels: }Wind velocity profile, in solid black for the 2D spatially averaged model of \citet{Debnath+2024}, dashed-red for the 1D PoWR model with the best-fit parameters from Table~\ref{tab:params} and $\varv_{\mathrm{turb}}=0$, and in solid red for the same parameters but $\varv_{\mathrm{turb}}>0$, from left to right for the O8, O4, O2 spectral types, respectively. \textit{Middle panels: }Same as the \textit{upper panel} but for the density profile. \textit{Lower panel:} Same as the \textit{upper panel} but for the gas temperature. }
      \label{fig:profile}
  \end{figure*}

In Fig.\,\ref{fig:profile}, we show the full set of wind velocity, density, and gas temperature profiles for the O8, O4 and O2 models, respectively. To isolate the influence of the turbulence, we now show only the lateral- and temporal-averaged 2D model stratifications by \citet[black]{Debnath+2024} together with the ``best-fit'' 1D model $\varv_{\mathrm{turb}}>0$ (red, solid). For comparison, we further show the stratifications of a model adopting all the ``best-fit'' values, but assuming $\varv_{\mathrm{turb}}=0$ (dashed red). An overview of the corresponding stellar parameter results are given in Table~\ref{tab:params}. The mass-weighted velocity and density profiles for the averaged 2D model is also shown in Fig.~\ref{fig:massvel}.

For convenience, all panels in Fig.\,\ref{fig:profile-O4} and \ref{fig:profile} are using the $x$-coordinate defined as
\begin{equation}\label{eq:x}
    x=1-R_{2/3}/r
\end{equation} to sufficiently highlight the region around the photospheric radius $R_{2/3}$. It is evident that in the absence of turbulence, there is a steep increase of the density in the inner and photospheric parts ($x \sim0$, Eq.~\ref{eq:x}), while the density gradient becomes more shallow with increasing turbulent pressure. We also plot the same parameter stratifications as in Fig.~\ref{fig:profile} but with respect to the Rosseland optical depth in Fig.~\ref{fig:profile-taur}.

When introducing a turbulent velocity $\varv_{\mathrm{turb}} > 0$ for the solution of Eq.~\ref{eq:as}, the value is varied such that the 2D averaged density profile is reproduced as accurate as possible. Our best-fit results are $\varv_{\mathrm{turb}}=35$, 88, and 106 km\,s$^{-1}$ for the O8, O4, and O2 stellar models, respectively. For the 2D models from \citet{Debnath+2024}, a resulting average turbulent velocity can be defined as the density-weighted average root mean square velocity
\begin{equation}
\overline{\varv_{\mathrm{turb}}}=\sqrt{\overline{P_{\mathrm{turb}}}/\langle\rho\rangle} \mbox{~\text{with}~} \overline{P_{\mathrm{turb}}} = \langle\rho\varv_r^2\rangle\text{.}    
\end{equation}
The resulting values are $\sim$$30$\,km\,s$^{-1}$ for the O8 model, $\sim$$60\dots80$\,km\,s$^{-1}$ for the O4, and $\sim$$100$\,km\,s$^{-1}$ for the O2. While this set is comparable to our 1D results, our values are systematically somewhat higher, in line with the 1D test calculation for the O4 model performed in \citet{Debnath+2024}, which needed $\sim$$90$\,km\,s$^{-1}$.

  \begin{table*}
\caption{Best fit final parameters for the 1D PoWR model to match the profile of \citet{Debnath+2024}. All fundamental parameters are scaled at $\tau=2/3$. We also include the fundamental parameters from Table 1 of \citet{Debnath+2024} for comparison. 
}
\label{tab:params}
\small
\centering
\begin{tabular}{c c c c c c c c c c c c c}
    \hline \hline
     M. & $\log(L_{\star}/L_{\odot})$ & $\log\dot{M}$ & $\log\dot{M}^{D+24}$ & $T_{\mathrm{eff}}$ & $T_{\mathrm{eff}}^{D+24}$ & $M_{\star}$ & $R_{\tau=2/3}$ & $R_{\tau=2/3}^{D+24}$ & $\log g$ & $\log g^{D+24}$ & $\varv_{\mathrm{turb}}$ & $\beta$ \\
      &  & ($M_{\odot}\,\mathrm{yr}^{-1}$) & ($M_{\odot}\,\mathrm{yr}^{-1}$) & (kK) & (kK) & ($M_{\odot}$) & ($R_{\odot}$) & ($R_{\odot}$) &  &  & (km\,s$^{-1}$) &  \\    
     \hline 
        O8 &  5.23 & -6.75 $\pm$ 0.25 & -6.86 & 33.1 & 33.3 & 26.9 & 12.56 & 12.26 & 3.67 & 3.69 & 35 $\pm$ 25 & 1.01 \\
        O4 &  5.78 & -5.55 $\pm$ 0.25 & -5.84 & 38.3 & 39.6 & 58.3 & 17.63 & 16.98 & 3.71 & 3.74 & 88 $\pm$ 50 & 1.01 \\
        O2 & 5.93 & -5.26 $\pm$ 0.25 & -5.56 & 40.9 & 43.8 & 58.3 & 18.39 & 15.99 & 3.67 & 3.79 & 106 $\pm$ 50 & 1.01 \\
     \hline
     
\end{tabular}

\end{table*}

In order to further improve the agreement of the 1D velocity and density profiles with the 2D averages, we change the initial parameters, {such as $\beta$ or $\dot{M}$, which directly impact $\rho(r)$ due to Eq.\,\eqref{eq:conteq}.} We vary $\log\dot{M}\,[M_{\odot}\,\mathrm{yr}^{-1}]$ in steps of $\Delta (\log\dot{M}\,[M_{\odot}\,\mathrm{yr}^{-1}]) = 0.25$ and $\varv_{\mathrm{turb}}$ [km\,s$^{-1}$] in steps of $\Delta \varv_{\mathrm{turb}}=25$ km\,s$^{-1}$ for the O8 model and 50 km\,s$^{-1}$ for the O4 and 02 models. 
In the resulting best-fit models for the 2D averaged profiles, several fundamental parameters change  compared to the values from \citet{Debnath+2024}: for the late O8 $\log\dot{M}\,[M_{\odot}\,\mathrm{yr}^{-1}]$ changes from $-6.86$ to $-6.75$, $T_{\mathrm{eff}}$ decreases by 200 K, and the corresponding radius increases by 0.3\,dex. For the O4 model, $\dot{M}$ increases by 0.29\,dex,  $T_{\mathrm{eff}}$ decreases by 1300\,K and the corresponding radius increases by 0.65 $R_{\odot}$.
For the O2 model, $\dot{M}$ increases by 0.3\,dex, $T_{\mathrm{eff}}$ decreases by 2900 K while the radius increases by 2.4\,$R_{\odot}$. The surface gravity for this model has the most dramatic change of 0.12\,dex while for the other models the change is $<0.03$ dex compared to the 2D models. The luminosities and stellar masses remain the same as \citet{Debnath+2024}. While in particular the changes in $T_\text{eff}$ might appear strange at first, they are a direct consequence of the increased mass-loss rates as $T_\text{eff} \equiv T_{2/3}$ is an output parameter of the PoWR models while the flux at the inner boundary (expressed as an effective temperature $T_\ast$) is kept fixed.

When comparing the photosperic radius and effective temperature of the 1D PoWR models to the averaged 2D RHD model values, it is important to consider how the averaged quantities are estimated in Table 1 of \citet{Debnath+2024}: First, $\tau=2/3$ is calculated in every single radial ray of their multi-dimensional simulations. Then, the unweighted average value is obtained for the photospheric radius and flux-temperature. This method leads to a significant spread in the values \citep[see Figures 14 and 15 in][]{Moens+2022wr} and could partially explain the different values needed to approximately match 1D and 2D approaches.

In summary, the differences in parameters compared to \citet{Debnath+2024} are 
%the most dramatic 
largest for the earliest model. The O2 model is also the most extreme model with the highest $\dot{M}=10^{-5.26}\,M_{\odot}\,\mathrm{yr}^{-1}$ and $\varv_{\mathrm{turb}}=106$ km\,s$^{-1}$. Potential implications of these differences are discussed in Section~\ref{sec:diss}.

As evident in particular from the upper panels of Fig.\,\ref{fig:profile-O4} and \ref{fig:profile}, the settings for the connection point also have a profound influence. In all models with $\varv_\mathrm{turb} > 0$, the velocity increase starts much further out, which is a consequence of including the turbulent velocity in the connection requirement, and needed to obtain the significantly shallower density profile suggested by the averaged 2D model. On the other hand, this then leads to a mis-match in velocities around the wind launching point, which now lies significantly further out in the 1D PowR model than in the averaged 2D simulation. In Section\,\ref{sub:vel} we investigate the effect of shifting the connection point.

\subsection{Spectral synthesis}

\begin{figure}
      \centering
      \includegraphics[width=.8\linewidth]{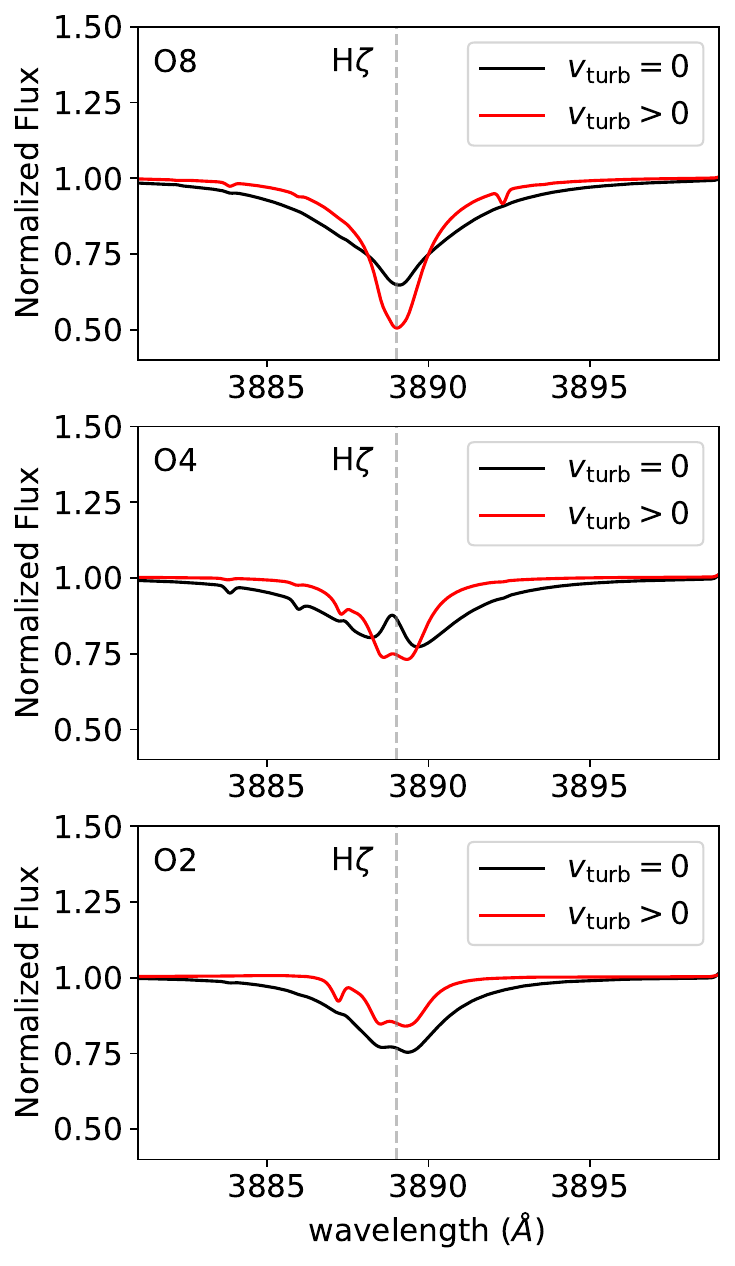}
      \caption{Normalized flux for the H$\zeta$ line from top to bottom for the PoWR model spectra calculated for the O8, O4, O2 models. The red spectrum shows the profile resulting from the model incorporating $\varv_{\mathrm{turb}}>0$ to reproduce the 2D average density profile. In black, the spectral lines from a PoWR models with the same parameters, but $\varv_{\mathrm{turb}}=0$ is shown. }
      \label{fig:linesh}
  \end{figure}

\begin{figure}
      \centering
     \includegraphics[width=.8\linewidth]{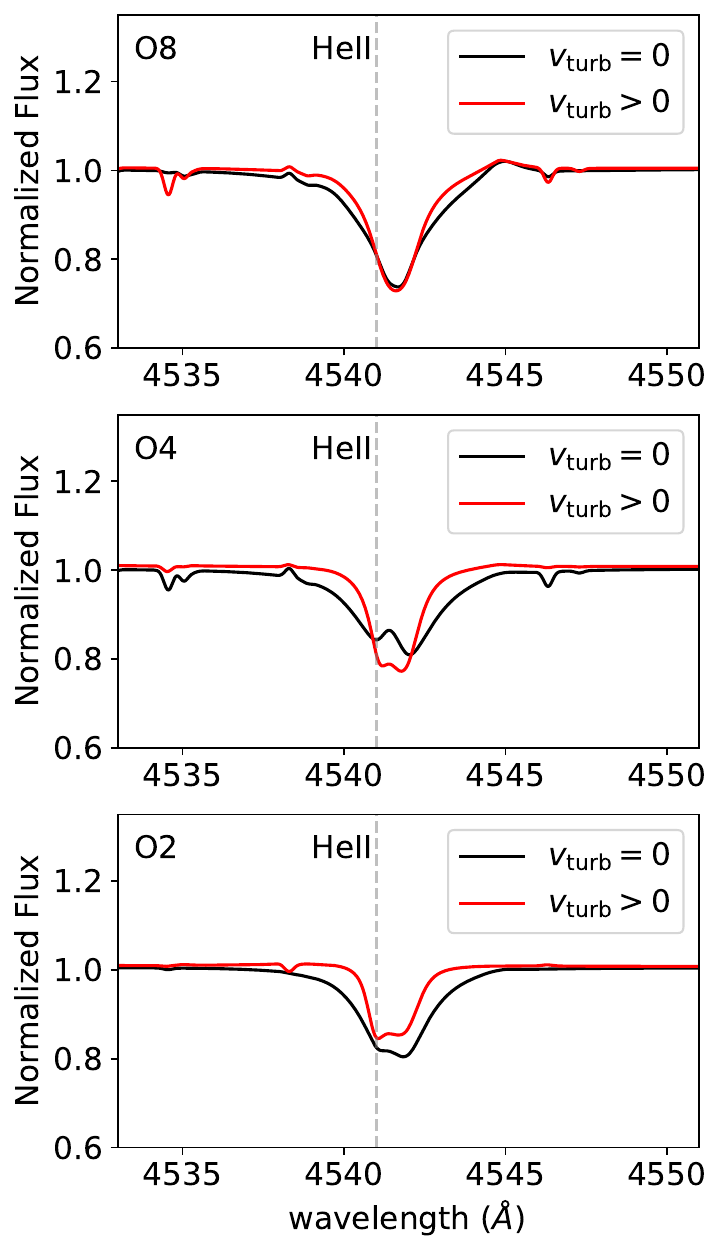}
      \caption{ Same as Figure~\ref{fig:linesh} but for the \ion{He}{ii} $ - 4541\,$\AA\ line.}
      \label{fig:lineshe}
  \end{figure}

\begin{figure}
      \centering
     \includegraphics[width=.85\linewidth]{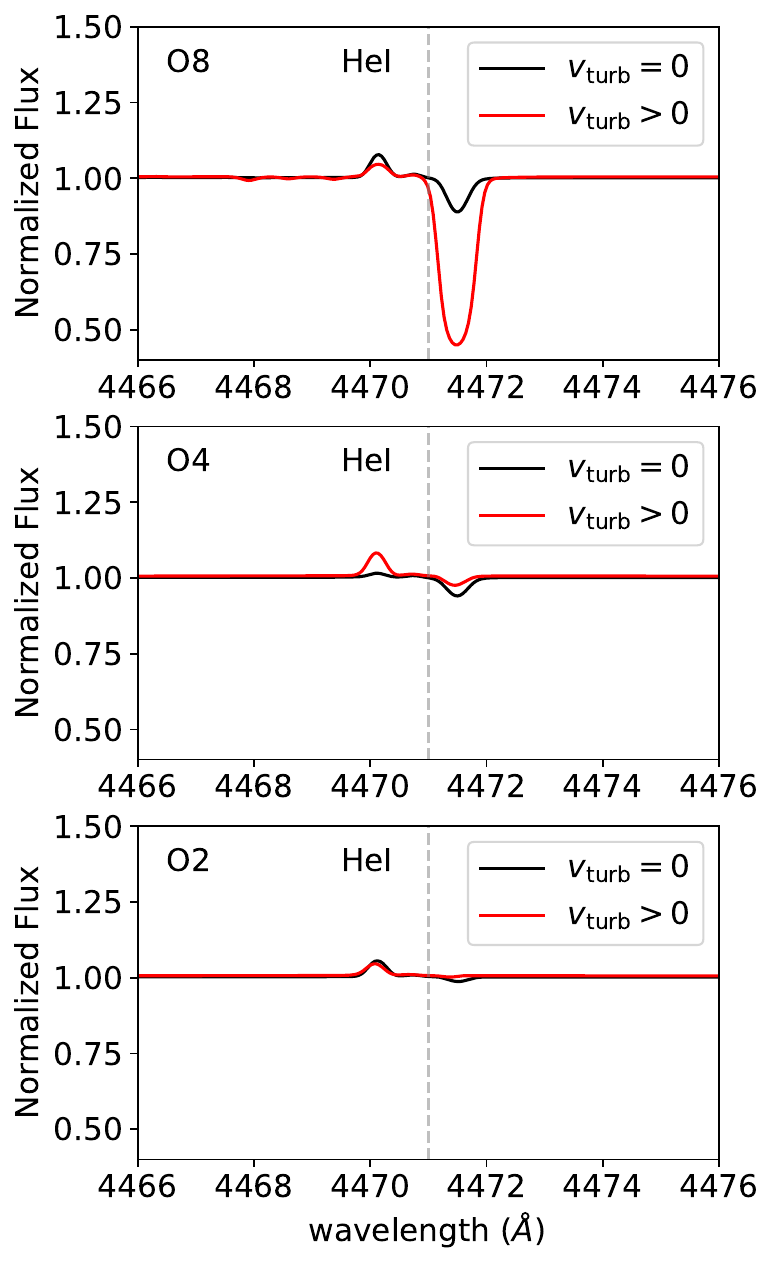}
      \caption{ Same as Figure~\ref{fig:linesh} but for the \ion{He}{i} $ - 4471\,$\AA\ line.}
      \label{fig:lineshei}
  \end{figure}

To study the implications of the structures predicted by \citet{Debnath+2024} on the spectral diagnosis, we predict the emergent spectra for all models using the corresponding 1D PoWR models.
PoWR calculates the emergent spectrum from a converged 1D model. The radiative transfer is solved in a fixed reference frame (``observer's frame'') to obtain the emergent flux and intensity. Electron scattering and line broadening effects are included. Since we only compare model spectra, we do not include any further macroturbulent broadening.
Due to the comparably high mass-loss rates -- in particular for the O4 and O2 models -- not only \ion{He}{ii}$\,4686\,$\AA\, but also multiple Balmer series members are considerably filled in with wind emission. This can blur the effects of the turbulent pressure on the absorption-line wings. We therefore focus our discussion on the H$\zeta - 3889\,$\AA\ ,\ion{He}{ii} $ - 4541\,$\AA\ and \ion{He}{i} $ - 4471\,$\AA\  spectral lines as shown in Figs.\,\ref{fig:linesh}, \ref{fig:lineshe} and \ref{fig:lineshei}. In red, we show the model approximating the 2D density structure including an adjusted $\varv_{\mathrm{turb}} > 0$. For comparison, we also show the spectrum of the corresponding model with the same parameters, but removing the turbulent pressure ($\varv_{\mathrm{turb}}=0$).
It is immediately evident that introducing a $\varv_{\mathrm{turb}}>0$ in the hydrostatic equation can have significant consequences on the shape of the spectral lines. In particular for the wings of the lines, the zero turbulence model shows broader lines than the non-zero solution for the case of H$\zeta - 3889\,$\AA\  and \ion{He}{ii} $ - 4541\,$\AA\ . The reason is the higher effective surface gravity in the case of zero turbulent pressure (black profiles in Figs.~\ref{fig:linesh} and  \ref{fig:lineshe}). For the O8 model, the presence of \ion{He}{i} $ - 4471\,$\AA\ in Fig.~\ref{fig:lineshei} is highly affected by the turbulence term, while for the earlier types the presence is almost negligible. As the turbulent pressure reduces the effective gravity, one would need to lower the actual surface gravity of the model to mimic the shape of the red profiles if $\varv_{\mathrm{turb}} = 0$ is assumed (see Section~\ref{sec:logg}). The difference in the equivalent width for \ion{He}{i} $ - 4471\,$\AA\ when comparing results from a change in the general He ionization structure between the two models that is likely induced by the different connection regime resulting from the inclusion of turbulence.

%-----------------------------------------------------------------
\section{Discussion}\label{sec:diss}

      \begin{figure}
      \centering
      \includegraphics[width=.8\linewidth]{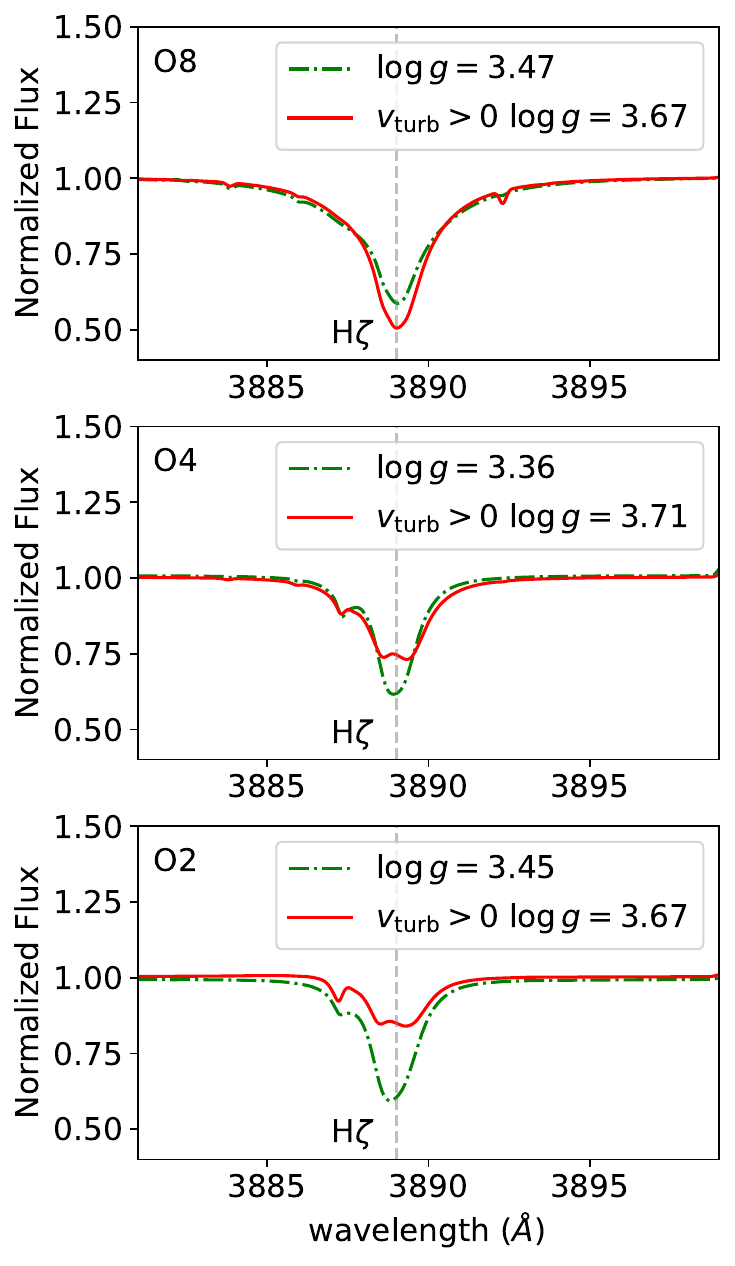}
      \caption{
      Normalized flux for the Hydrogen H$\zeta$ line from top to bottom for the O8, O4 and O2 stars. In solid red for the best-fit 1D PoWR model with $\varv_{\mathrm{turb}}>0$, in dashed green for $\varv_{\mathrm{turb}}=0$ and lower surface gravity. The stellar parameters for the different models are shown in Table~\ref{tab:mass}}.
      \label{fig:linesh2}
  \end{figure}

        \begin{figure}
      \centering
      \includegraphics[width=.8\linewidth]{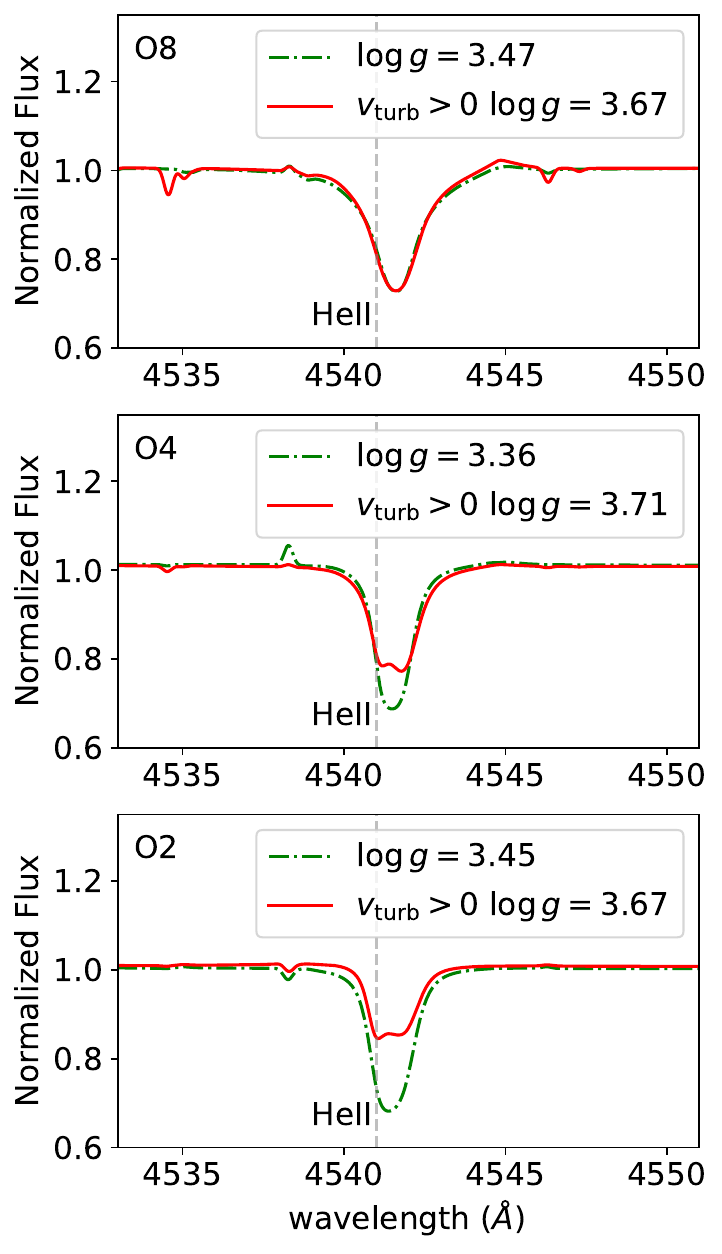}
      \caption{ Same as Figure~\ref{fig:linesh2} but for the He{\sc ii} line. 
      }
      \label{fig:lineshe2}
  \end{figure}

   \begin{figure}
      \centering
      \includegraphics[width=.85\linewidth]{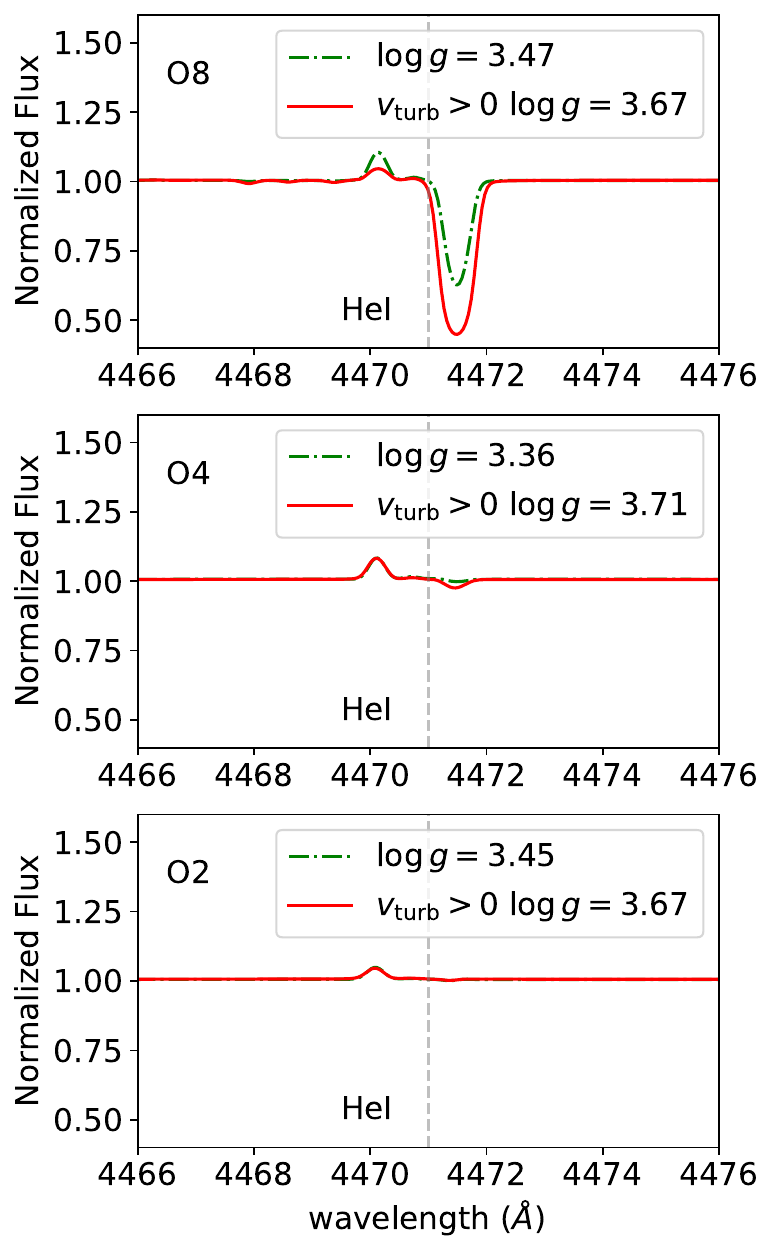}
     \caption{ Same as Figure~\ref{fig:linesh2} but for the He{\sc i} line. 
      }
      \label{fig:lineshei2}
  \end{figure}

\subsection{Turbulence and density profile}

We obtained a best-fit result of $\varv_{\mathrm{turb}}=35$, 88 and 106 km\,s$^{-1}$ for the O8, O4 and O2 stellar models, respectively. We notice that the turbulent velocity increases for earlier spectral types. This effect is a consequence of the closer proximity to Eddington Limit in the earlier models as the hotter 2D models from \citet{Debnath+2024} do not just have different effective temperatures, but also different stellar parameter combinations. As a consequence, the local excess of the Eddington limit at the depth of the iron opacity peak will be larger for earlier-type models, providing the energy and momentum for larger turbulent motion. This qualitatively aligns with earlier trend predictions from 1D stellar structure calculations for sub-surface convection caused by the hot iron-bump \citep[e.g.,][]{Cantiello+2009}.

In \citet{Debnath+2024}, the determined mean turbulent velocity is depth-dependent, i.e., $\varv_{\mathrm{turb}}(r)$, with a maximum value of $\sim$$80$\,km\,s$^{-1}$ for their O4 model in the sub-photospheric regime. Inwards, $\varv_{\mathrm{turb}}$ gradually decreases to zero in the layers below the iron bump. Outwards, the average turbulent pressure decreases, but the mean turbulent velocity generally increases, albeit not monotonically \citep[cf.\ Fig.\,16 in][]{Debnath+2024}. 

For our 1D PoWR models, already the assumption of a constant turbulent velocity is sufficient to reproduce the density profile in the outer layers ($x = 1-R_{2/3}/r >0$). 
The models presented in this work do not cover the regime below the iron opacity peak. Hence, the inner decrease of the turbulent velocity does not need to be covered and a constant value is sufficient (see Section~\ref{sec:2d}). However, in the middle panels of Fig.\,\ref{fig:profile}, as we go inwards (sub-photosperic layers where $x= 1-R_{2/3}/r<0$) the 1D $\varv_{\mathrm{turb}}>0$ profile stars differing from the averaged 2D, challenging our assumption of a constant $\varv_{\mathrm{turb}}$. 
Still, assuming a constant $\varv_{\mathrm{turb}}$ already significantly improves our fit and impacts the spectral lines. For future work, we plan to introduce a depth dependence on the turbulence.

\subsection{Wind velocity stratification and Mass-loss rate}\label{sub:vel}

Fig.\,\ref{fig:profile-O4} shows two different radial wind velocity profiles from the 1D models (upper panel): our initial approach, using the same basic stellar parameters as the 2D RHD model, employs a velocity law with $\beta=0.8$, motivated by \citet{Pauldrach+1986} as a solid blue line. In the second approach, we increase $\beta>1$ to $1.01$ (solid red profile in Fig.\,\ref{fig:profile-O4}). 
Fig.\,\ref{fig:profile} shows both models with $\beta=1.01$ with and without turbulence. For the wind velocity profile, the effect of the turbulence is to create an onset further out to $x\sim0$ (Eq.~\ref{eq:x}). Interestingly, this is not the case when comparing to the mass-weighted averages (cf.\ Fig.~\ref{fig:massvel}), where the onset of the averaged 2D models align much more with the onset we get for the 1D model with turbulence. If we would prescribe the mass-weighted $\varv(r)$ in a 1D model to infer the density via Eq.~\ref{eq:conteq}, one might now in turn expert offsets in the density. As shown in Fig.~\ref{fig:massvel}, the inner part is is problematic due to the complexity going on, but the wind regime looks more promising, although there is indeed a shift for the O8 and O2 model. As we focus on the turbulent (sub-)photospheric structures in this work, we continue our comparison with the non-weighted average.

In order to improve the fit for the radial velocity in Fig.~\ref{fig:profile}, we have changed the connection settings between the (quasi-)hydrostatic regime and the $\beta-$law. Fig.\,\ref{fig:vsonic} shows the velocity profile modifying the connection point to 0.95 the total local sound speed, instead of the initial 0.95 the effective sound speed accounting for the turbulence. Albeit the wind velocity profile matches the averaged 2D better, the shift on the connection point negatively affects the agreement with respect to the density around the photosphere.

Each of the spatially averaged 2D wind velocity profiles for the three models (solid-black profiles in the upper panels of Fig.\,\ref{fig:profile}) shows a deceleration region ($\varv_r < 0$) around or just below the photosphere ($x\sim0$, Eq.~\ref{eq:x}). This reflects that any deeper acceleration of pockets of gas is not sufficient to launch a steady wind, but instead create radiatively-driven turbulent motion and local regions of infalling gas as reported by \citet{Debnath+2024}.  
This non-monotonic behaviour in the (averaged) velocity field cannot be directly mapped in our 1D models where we connect the hydrostatic solution to a $\beta$-law. Testing the spectral imprint of a deceleration region in the photosphere goes beyond the current capabilities of 1D atmosphere models. Yet, the sub-surface impact of changing the pressure scale height is already accounted for with the introduction of the turbulent pressure term.

In all 2D models, the velocity increase eventually gets more shallow in the outer part, most prominently seen in the O8 model. Since we are assuming a single $\beta-$law for the 1D PoWR approach, we cannot reproduce this flattening observed in the 2D models. In principle, a more structured velocity law (e.g., a double-$\beta$-law) could be introduced to mitigate this issue. However, this would introduce even more free parameters, which are hard to constraining systematically given the small amount of available 2D models. 

As reported in Section\,\ref{sec:comp}, the mass-loss rates of our best-fit 1D models in comparison to the 2D models are increased by 0.1\,dex for the O8 model and 0.3\,dex for the earlier stellar types. As shown in Fig.\,\ref{fig:profile-O4}, this increase in $\dot{M}$ is necessary to yield the higher wind density from the 2D average density profile. The mismatch for the model with the same $\dot{M}$ could arise from trying to match structures obtained using different methodologies: For example, because of the strong anti-correlation between wind velocity and density in multi-D models \citep[][]{Moens+2022wr, Debnath+2024}, the average velocity profile $\langle\varv\rangle$ will be greater than the density averaged velocity, $\langle\rho\varv\rangle/\langle\rho\rangle$. Therefore, when trying to match a higher $\langle\varv\rangle$-structure, a somewhat higher $\dot{M}$ will be necessary to obtain the same $\langle\rho\rangle$. Future investigations of different averages will be needed to see what is best suited to map the structures from a 2D framework in 1D while preserving the averaged parameters as much as possible. Still, the obtained differences in $\dot{M}$ of a factor of two for the earlier models are not drastic, but would be noticeable in the spectral predictions.

\subsection{Electron temperature stratification}

While the wind velocity and density profiles can be relatively well reproduced by a PoWR model when compared to the averaged 2D profile (except for the velocity onset of the wind, see discussion in Section\,\ref{sub:vel}), the temperature stratification shows large differences between both modelling approaches (see the bottom panels in Fig.\,\ref{fig:profile}). For all models, the 2D temperature stratification is much smoother than those of the 1D models. The lower panels in Figs\,\ref{fig:profile-O4} and \ref{fig:profile} show the gas temperature stratification for both modelling approaches. The flux temperature calculated from the total emergent flux and the radiation temperature calculated from the integrated mean intensity for the PoWR model is shown in Fig.\,\ref{fig:trad}.

The main difference in the gas temperature profiles between the averaged 2D model and the 1D approach is that while the 2D temperature smoothly decreases, the 1D models present bumps where there is a temperature inversion in the wind regime, for both models with and without turbulence. For the case of the O8 star, there is a moderate temperature increase near the photosphere and the temperature decreases in the outer layers. Compared to the averaged 2D profile, the temperature in the wind region is higher. For the O4 and O2 cases, the temperature inversion occurs at outer layers of the atmosphere. The temperature is also significantly lower than the 2D averaged profile, in particular for the O2 model. Any inclusion of (micro-)clumping in the 1D models has little effect on the results. Fig.\,\ref{fig:desconprofile} comparing a smooth with a clumped 1D model reveals that the temperature profile, at least for the given mass-loss rate, is not significantly affected by the change in clumping properties.

These differences in the temperature profiles for both approaches are therefore more likely to arise from differences in the methods used to compute the energy balance in the 1D and 2D models. In the 2D models,
the frequency-integrated heating and cooling terms rely on frequency-integrated opacities that have been weighted with the relevant quantities: the energy mean opacity and the Planck mean opacity. In the optically thick static diffusion regime, both the energy- and Planck mean opacities, as well as the flux mean opacity present in the radiation force term are adequately described by the Rosseland mean. However, in the outer wind, where line driving and the Sobolev effect are important \citep[see][]{Sobolev1960}, the Rosseland mean is no longer an accurate approximation. Instead, the three different weighted opacities have to be calculated separately. As mentioned in Section\,\ref{sec:2d}, methods to get accurate flux-mean opacities have been developed by \citet{Poniatowski+2022}, but methods for the energy and Planck mean opacities are still work in progress. Instead, the multi-dimensional models approximate the energy and Planck mean by the flux mean. This very likely results in an over-efficient heating and cooling, which in practice forces the gas and radiation temperatures to the same value, the net effect being a higher gas temperature. This is in contrast to non-LTE models, which 
typically finds the radiation temperature outside of the photosphere to be higher than the gas temperature. On the other hand, in 1D models cooling due to adiabatic expansion in the wind is often neglected and the temperature structure rather obtained from assuming strict radiative equilibrium (or via the equivalent electron thermal balance formalism). It thus remains an open question whether the characteristic temperature inversion often found in 1D atmospheric models \citep[see also, e.g., discussion in][]{Puls+2005} would survive in multi-D models with a better treatment of the wind energy balance.

\subsection{Surface gravity}\label{sec:logg}

If the non-zero turbulence ($\varv_{\mathrm{turb}}$ > 0) is included in Eq.\,\eqref{eq:as}, it changes the effective pressure scale height in the (quasi-)hydrostatic regime and therefore affects the wings of pressure-broadened absorption lines. Thereby, the fundamental diagnostics for the surface gravity $\log g$ are affected such that for $\varv_{\mathrm{turb}}$ > 0 a larger $\log g$ is required to approximately get the same line wings as for $\varv_{\mathrm{turb}}$ = 0. For an assumed constant $T(r) \equiv T_\text{phot}$ in the relevant (quasi-)hydrostatic regime, the difference can be estimated as 

\begin{equation}\label{eq:logg}
    \Delta (\log g)\approx\log \left( 1+\frac{\varv_{\mathrm{turb}}^{2}\mu m_{\mathrm{H}}}{ k_{\mathrm{B}}T_\text{phot}} \right) \equiv \log \left( 1+\frac{\varv_{\mathrm{turb}}^{2}}{ a_\text{s}^2(T_\text{phot})} \right)\text{.} 
\end{equation}

To illustrate and check this difference in $\log g$, we have decreased the $\log g$ for our models with the parameters in Table~\ref{tab:params} and without including any turbulence ($\varv_{\mathrm{turb}}=0$), denoted as "$\log g_{0}$" in Table~\ref{tab:mass}. The profile comparison for the "$\log g_{0}$"  models is shown in Fig.\,\ref{fig:profile2}. We compare the spectral lines with the two models in Figs.\,\ref{fig:linesh2} and \ref{fig:lineshe2}: in dashed-green the model without any turbulence and decreased $\log g_{0}$ and in solid-red the model with ($\varv_{\mathrm{turb}}>0$) and higher $\log g$. We notice that for the O2 PoWR model, we could not reach lower $\log g_{0}$ because of convergence problems, so the results shown in Table~\ref{tab:mass} for O2 are upper limits. We also take into account the correlation between derived $\log g$ and $T_{\mathrm{eff}}$ leaving $T_{\mathrm{eff}}$ as a free parameter in the models, so that a higher $\log g$ requires a higher $T_{\mathrm{eff}}$ to keep the He ionization balance correct.

We estimate the surface gravity difference using Eq.~\ref{eq:logg}, assuming $\mu\sim0.6$ as of a fully ionized plasma and $T_\text{phot}=T_{\mathrm{eff}}$. The difference that we calculate is $\Delta (\log g)= 0.4$, 0.9, 1.0, for the O8, O4 and O2 models, respectively. When compared to $\log g_{0}$ in Table~\ref{tab:mass}, we obtain a $\Delta (\log g)=0.2$, $0.4$ and $>0.2$ instead.
The difference in $\log\,g$ estimated with Eq.\,\ref{eq:logg} differs from the difference when using the values from Table\,\ref{tab:mass} by a factor of two. On the one hand, the differences in fitting the wings of the lines can give rise to a lower $\Delta (\log g_{0})$ than predicted by Eq.\,\ref{eq:logg}. On the other hand, a secondary effect would be the feedback from the Thomson radiative acceleration for the model without turbulence, $\Gamma_{\text{e},0}$, as the mass changes but the luminosity remains constant. The $g_{\mathrm{turb}}$ can be estimated as 
\begin{equation}\label{eq:logggamma}
g_{\mathrm{turb}} \approx g_0 \left( 1+\frac{\varv_{\mathrm{turb}}^{2}}{a_\text{s}^2(T_\text{phot})} \right) 
\frac{1-\Gamma_{\text{e},0}}{1- \Gamma_{\text{e},0} g_0/g_{\mathrm{turb}}}\text{,}
\end{equation}
where $g_{\mathrm{turb}}$ is the surface gravity for the model with $\varv_{\mathrm{turb}}>0$ and $g_{0}$ for the model without turbulence, and $\Gamma_{e,0}$. Using Eq.\,\ref{eq:logggamma}, we find a $\Delta (\log g)= 0.12$, 0.75, 0.81 for the O8, O4 and O2 models, respectively. For the O8 model, the inferred difference from Eq.\,\ref{eq:logggamma} is 0.08 dex smaller than with the surface gravities from Table\,\ref{tab:mass}, providing a closer estimate than with Eq.\,\ref{eq:logg}. The large value obtained in Eq.\,\ref{eq:logggamma} for the O2 model is above the lower limits differences inferred from the surface gravities in Table\,\ref{tab:mass}. Eqs.~\ref{eq:logg} and \ref{eq:logggamma} do not take into account the $\log g$ and $T_{\mathrm{eff}}$ correlation mentioned earlier, which could explain the difference in the estimated $\Delta (\log g)$ inferred from the models.

 Figs.\,\ref{fig:linesh2} and \ref{fig:lineshe2} show the effect of lowering the surface gravity for a model without turbulence as compared to a model with $\varv_{\mathrm{turb}}>0$ and higher surface gravity in the spectral lines H$\zeta - 3889\,$\AA\ and \ion{He}{ii} $ - 4541\,$\AA\, where we are able to fit the width of the wings with both the $\varv_{\mathrm{turb}}>0$ and $\log g$ model and the $\varv_{\mathrm{turb}}=0$ and $\log g_{0}$ model. The effect is particularly visible for the O8 and O4 models. While for the O2  model we still cannot perfectly reproduce the width of the wings due to our results being upper limits, the effects are already visible. Both $\log g$ and $\log g_{0}$ models infer two different spectroscopic masses, shown in Table~\ref{tab:mass} as $M_{\mathrm{turb}}/M_{\odot}$ and $M_{0}/M_{\odot}$ for the model with turbulence and without turbulence, respectively.
 Fig.~\ref{fig:lineshei2} compares both the $\varv_{\mathrm{turb}}>0$ and $\log g$ model and the $\varv_{\mathrm{turb}}=0$ and $\log g_{0}$ model for \ion{He}{i} $ - 4471\,$\AA\ . In this case, the width of the wings is relatively well reproduced by the O8 model, while the lines are very faint in the O4 and O2 models, but still yield a similar shape for both the $\varv_{\mathrm{turb}}>0$ and $\log g$ model and the $\varv_{\mathrm{turb}}=0$ and $\log g_{0}$ model.

To study the effect on the spectral lines caused by the microturbulence broadening term, we have modified $\xi_{\mathrm{min}}$. Fig.\,\ref{fig:xi} shows the H$\zeta - 3889\,$\AA\ line for the O8 model, both for the $\varv_{\mathrm{turb}}>0$ and $\log g$ model and the lower surface gravity model $\log g_{0}$ and $\varv_{\mathrm{turb}}=0$. Moreover, we have included different $\xi_{\mathrm{min}}$ in the non-turbulent model to see if we could fit the $\varv_{\mathrm{turb}}>0$ line, but we were unable to fit perfectly the depth of the line. Therefore, the influence of $\xi_{\mathrm{min}}$ cannot explain the effect on the $\varv_{\mathrm{turb}}>0$ for the depth of the lines.

The analysis of O stars have so far neglected the addition of a turbulence term in the solution of the hydrostatic equation. However, for B supergiants \citet{Wessmayer+2022} performed a quantitative spectroscopic analysis using the ATLAS12 code \citep{Kurucz2005} which allows for the inclusion of turbulent pressure. The turbulence was limited to a maximum of $14\,\mathrm{km\,s}^{-1}$ with resulting $\Delta(\log g) \approx 0.05\,$dex. \citet{Wessmayer+2022} show that the addition of a  modest turbulence can affect the lines of helium and some metallic lines (\ion{Si}{ii}, \ion{C}{ii} and \ion{S}{ii}).

\subsection{Mass discrepancy}

Without the turbulence term but decreased surface gravity, we obtain masses that are $\sim$$10\,M_\odot$, $\sim30\,M_\odot$ and $>25\,M_\odot$ lower than when accounting for the turbulence, for the O8, O4 and O2 models, respectively, while the photospheric radius changes by $<0.01\,R_{\odot}$. Noticing the significant change in stellar masses, it is interesting to point out that adding a turbulence term in the hydrostatic equation could therefore potentially solve the so-called ``mass discrepancy'' problem.

  \begin{table*}
\caption{Surface gravities, $T_{\mathrm{eff}}$ and masses for $\varv_{\mathrm{turb}}>0$, $\log\,g_{\mathrm{turb}}$ and $M_{\mathrm{turb}}/M_{\odot}$, and for $\varv_{\mathrm{turb}}=0$, $\log\,g_{0}$ and $M_{0}/M_{\odot}$ for the three different O star models, as well as the evolutionary mass determined from the HRD position.}
\label{tab:mass}
\small
\centering
\begin{tabular}{c c c c c c c c}
    \hline \hline
     Model & $\log\,g_{\mathrm{turb}}$ & $T_{\mathrm{eff-turb}}$ (kK) & $M_{\mathrm{turb}}/M_{\odot}$ & $\log\,g_{0}$ & $T_{\mathrm{eff-0}}$ (kK) & $M_{0}/M_{\odot}$ & $M_{\mathrm{evol}}/M_{\odot}$ \\
     \hline 
        O8 &  3.67 & 33.1 & 26.9 & 3.47 & 33.0 & 17.17 & 27.5 \\
        O4 &  3.71 & 38.3 & 58.3 & 3.30 & 37.4 & 25.23 & 51.5\\
        O2 &  3.67 & 40.9 & 58.3 & $<$3.45 & $<$40.6 & $<$33.60 & 64.3 \\
     \hline
     
\end{tabular}
\end{table*}

\begin{figure}
      \centering
  \includegraphics[width=1.\linewidth]{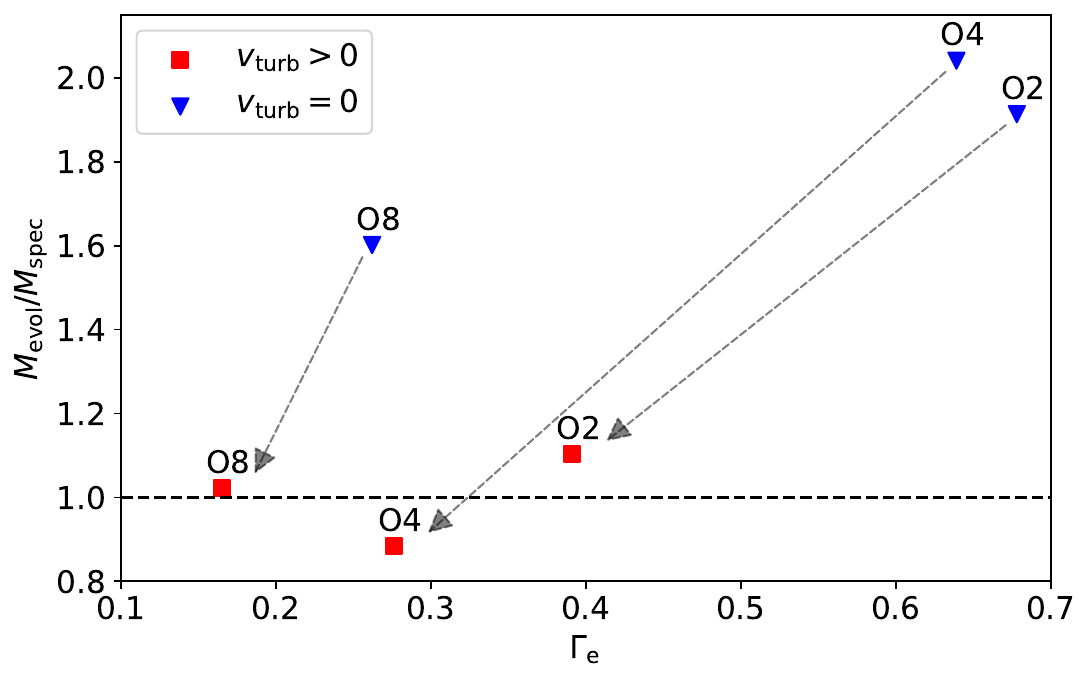}
  \caption{Ratio of $M_{\mathrm{evol}}/M_{\mathrm{spec}}$ with respect to the Thomson radiative acceleration $\Gamma_{e}$ for the models without turbulence (blue triangles) and with turbulence (red squares). We see a shift when we include a turbulent term in the models, obtaining a $M_{\mathrm{evol}}/M_{\mathrm{spec}}$ ratio closer to one.}
  \label{fig:mevolgamma}
\end{figure}

The ``mass discrepancy'' \citep[e.g.,][]{Herrero+1992} for hot stars refers to a difference between stellar mass estimations from evolutionary tracks and the positions of the stars in the Hertzsprung-Russell diagram (``evolutionary masses'') and the spectroscopically derived masses. Typically, the evolutionary masses are higher than the spectroscopic masses. In the original report of the problem by \citet{Herrero+1992}, stars closer to the Eddington limit showed stronger discrepancy, which would align very well with the obtained turbulence trend. 
Comparing the luminosities and effective temperatures of our models (see Table\,\ref{tab:params}) with the stellar evolution tracks from \citet{Ekstroem+2012}, we estimated the corresponding evolutionary masses, $M_{\mathrm{evol}}/M_{\odot}$ (see Appendix~\ref{app:evolmasses} for details), shown in Table~\ref{tab:mass}. Fig.\,\ref{fig:mevolgamma} shows the $M_{\mathrm{evol}}/M_{\mathrm{spec}}$ ratio with respect to the Thomson radiative acceleration $\Gamma_{\mathrm{e}}$ for the models with $\varv_{\mathrm{turb}}>0$ (red squares) and the models with $\varv_{\mathrm{turb}}=0$ and $\log\,g_{0}$ (blue triangles). We see a clear shift when we include a turbulent term in the models, obtaining a $M_{\mathrm{evol}}/M_{\mathrm{spec}}$ ratio closer to one, and therefore reconciling both mass determinations.

When comparing to spectroscopic data, \citet{Markova+2018} analysed optical spectra of 53 Galactic O stars using the FASTWIND and CMFGEN codes and compared to published predictions of evolutionary model grids by \citet{Brott+2011} and \citet{Ekstroem+2012}. We have checked an example star HD 46223 from \citet{Markova+2018} with spectral type O4 V and comparable stellar parameters to our O4 model such as the effective temperature ($T_{\mathrm{eff}}=43.5\pm1.5$ kK), surface gravity ($\log\,g=3.95\pm0.1$), stellar radius ($R_{\star}=10.9\pm2.0\,R_{\odot}$) and luminosity ($\log(L_{\star}/L_{\odot})=5.58\pm0.17$), while they obtain a spectroscopic mass ($M_{\mathrm{spec}}=38.9\pm14.4\,M_{\odot}$) which is $\sim$$20\,M_{\odot}$ lower than our estimated stellar mass for the O4 model including the turbulence term. 
 For the case of HD 46223 in \citet{Markova+2018}, all the evolutionary track predictions overestimate the mass by at least $8\,M_{\odot}$, obtaining an $M_{\mathrm{evol}}$ comparable to ours when including the turbulence ($M_{\mathrm{evol}}=47.0-51.8\,M_{\odot}$, depending on the evolutionary track used). In this particular example, we find that one systematically underestimates the spectroscopic mass, and accounting for the turbulent pressure in the hydrostatic equation could potentially reconcile both mass measurements. However, we notice that the $\log\,g$ for HD 46223 is significantly higher than for our O4 model. Therefore, a need for more spectroscopic studies including a wider range of spectral types and a turbulence term would be needed to determine the regime where the mass discrepancy could be reconciled.

\subsection{Influence on the radiative force}

\begin{figure}
      \centering
  \includegraphics[width=1.\linewidth]{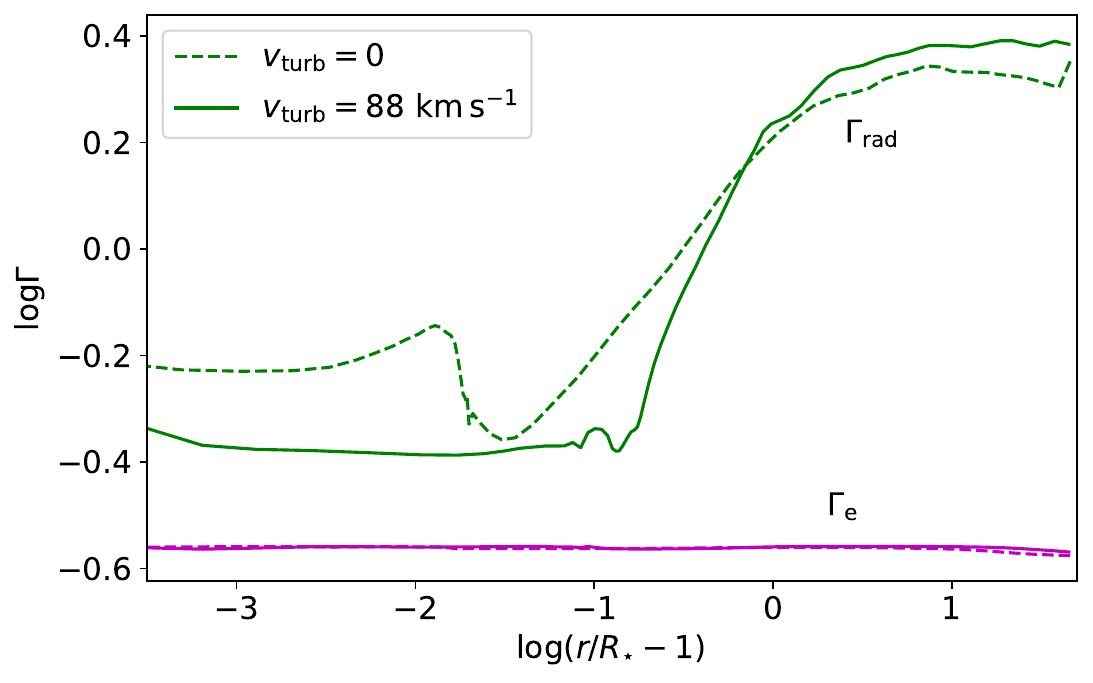}
  \caption{Radiative acceleration (total and Thomson) normalized to gravity as a function of radius for two 1D models resembling the O4 2D average. The two models only differ in their turbulent pressure which is included in the (quasi-)hydrostatic domain.}
  \label{fig:gamrad}
\end{figure}

In the 1D PoWR models, the inclusion of turbulent pressure in the (quasi-)hydrostatic regime also has a profound impact in the resulting radiative acceleration of the converged stationary models. In Fig.~\ref{fig:gamrad}, we plot the radiative acceleration normalized to gravity ($\Gamma_\text{rad}$) for our 1D models with and without turbulent pressure approximating the 2D-average from the O4 model by \citet{Debnath+2024}. The characteristic ``dip'' in the radiative acceleration that typically occurs just before the onset of the wind \citep[see, e.g.,][]{Sander+2015,Sundqvist+2019} is only prominent in the model without turbulent pressure, while the model including turbulence has a much smoother -- albeit lower -- radiative acceleration in the hydrostatic regime.
For comparison, we further show the radiative acceleration for the O8 and O2 1D models in Appendix~\ref{app:acc}. Similarly to the O4 model, the ``dip'' in the O2 case is smoothed out when introducing a $\varv_{\mathrm{turb}}$ term. This is not the case for the O8 model, where the ``dip'' is changed, but remains prominent. The reason is the more modest value of $\varv_{\mathrm{turb}}=35$ km\,s$^{-1}$ in comparison with the earlier O-types, which seems to be insufficient to avoid the formation of a pronounced ``dip''.  

Our 1D models in this work are not hydrodynamically-consistent and therefore cannot predict mass-loss rates intrinsically. Yet, they are sufficient to conclude that the inclusion of a turbulent pressure could have significant consequences on mass-loss predictions due to the resulting change of the $\Gamma_\text{rad}$-slope. 
The lower average of $\Gamma_\text{rad}$ in the (quasi-)hydrostatic regime could give rise to the assumption that resulting mass-loss rates from 1D dynamical calculations might be even lower. On the other hand, the avoidance of the ``dip'' is expected to change the iteration process and might avoid reaching much lower values, at least in regimes where considerable turbulent velocity is expected and provides an additional outward-directed force. With the presumed trend towards higher turbulence for earlier spectral types, the consequences are expected to be larger for hotter and more massive stars. None of this is so far covered in any of the existing theoretical mass-loss prescriptions for OB stars \citep[e.g.,][]{Vink+2001, KrtickaKubat2018, Bjoerklund+2023} with the resulting effects likely being very regime-specific, potentially leading to stronger differences between supergiants -- which are closer to the Eddington limit -- and dwarfs. On the other hand, the average mass-loss rates computed from the three 2D O-star simulations investigated here align fairly well with the predictions by \citet{KrtickaKubat2018, Bjoerklund+2023} \citep[see Fig.\,18 in][]{Debnath+2024}, which is surprising given the lack of turbulent pressure in the 1D models underlying these $\dot{M}$-predictions. New dynamically-consistent 1D calculations will be necessary to investigate this further, but are beyond the scope of the present exploratory paper.

\section{Conclusions}\label{sec:conc}
We calculated 1D stellar atmosphere models with the PoWR code to resemble the averaged 2D profiles of three O-star RHD simulations from \citet{Debnath+2024}. 
The average density profiles can be well reproduced by including a turbulent velocity term in the solution of the hydrostatic equation of motion. While \citet{Debnath+2024} find a depth-dependency for their inferred turbulent velocity, a constant turbulent velocity is sufficient to reproduce the relevant part of the mean 2D density profile in our 1D models reasonably well. This simplification is possible due to the fact that in the 2D models a large increase of the averaged turbulence happens in the optically thick deeper layers which do not influence the spectrum. In the future, we plan to include a depth-dependent turbulence to test variations of the turbulent velocity, its predicted increase in the supersonic regime and the potential impact on the inferred wind driving. 

In our attempt to reproduce the wind velocity profile, we employed a $\beta=1.01-$law for the supersonic part of the model. While this differs from the most-common value for O-stars of $\beta=0.8$ \citep[e.g.,][]{Pauldrach+1986}, increasing $\beta$ improves the connection to the (quasi-)hydrostatic part and shifts the increase in the velocity outwards, thereby producing a similar slope to the profiles from \citet{Debnath+2024}. A closer connection point between the hydrostatic and the $\beta$-law domain shifts the increase of the velocity inwards, thereby improving the agreement with the 2D-average velocity field, but at the same time worsens the fit of the 2D-averaged density distribution in the transition regime. It thus remains a challenge to find suitable velocity laws for this important transition region in unified model atmospheres connecting a quasi-hydrostatic photosphere to an outflowing wind described by an analytic $\beta$-law. Considering weighted averages such as the mass-weighted average might help to reconcile discrepancies in the wind-onset and outer wind region. Ideally, this should be investigated jointly with observations and diagnostics sensitive to this region, which is beyond the present prototype study. 

The addition of a turbulent pressure also has an impact on the derived radiative acceleration in the (quasi-)hydrostatic domain of the 1D PoWR models. When a significant turbulent motion is added in the solution of the hydrostatic equation, the acceleration slope becomes smoother and typical ``dip''-like structures are avoided. With its effect on the radiative acceleration profile, the inclusion of turbulent pressure will most likely also affect future mass-loss rate predictions from 1D hydrodynamically-consistent model atmospheres. 

The change of the scale height by the introduction of a non-negligible turbulent pressure term affects the resulting spectral line diagnostics, yielding higher values for $\log g$ and $M_{\mathrm{spec}}$ compared to similar models not including turbulence. Therefore, including this extra turbulent pressure could diminish or in some cases even remove the so-called ``mass discrepancy'' between inferred evolutionary and spectroscopic masses. Eventually, our findings underline the need for a better treatment of the outer layers in evolutionary models to enable more coherent comparisons with spectroscopic results. While we have discussed only three sets of models in this work, a broader investigation of the O and B star regime with both multi-D and 1D models will be necessary to infer proper associations of $\varv_{\mathrm{turb}}$ with spectral types, macroturbulence, or other parameter correlations necessary to systematically account for turbulent pressure in the spectral analysis of massive stars.

\begin{acknowledgements}
GGT is supported by the German Deutsche Forschungsgemeinschaft (DFG) under Project-ID 496854903 (SA4064/2-1, PI Sander). 
AACS, RRL, and JJ are supported by the German Deutsche Forschungsgemeinschaft (DFG) under Project-ID 445674056 (Emmy Noether Research Group SA4064/1-1, PI Sander). GGT and AACS further acknowledge support from the Federal Ministry of Education and Research (BMBF) and the Baden-W{\"u}rttemberg Ministry of Science as part of the Excellence Strategy of the German Federal and State Governments. JOS, DD, LD, NM, CVdS acknowledge the support of the European Research Council (ERC) Horizon Europe grant under grant agreement number 101044048 (ERC-2021-COG, SUPERSTARS-3D). OV, JOS, and LD acknowledge the support of the Belgian Research Foundation Flanders (FWO) Odysseus program under grant number G0H9218N and FWO grant G077822N.
\end{acknowledgements}

   \bibliographystyle{aa} % style aa.bst
   \bibliography{sample} % your references Yourfile.bib

\begin{appendix}

\section{O4 smooth model}\label{app:clump}
      \begin{figure}
      \centering
      \includegraphics[width=.85\linewidth]{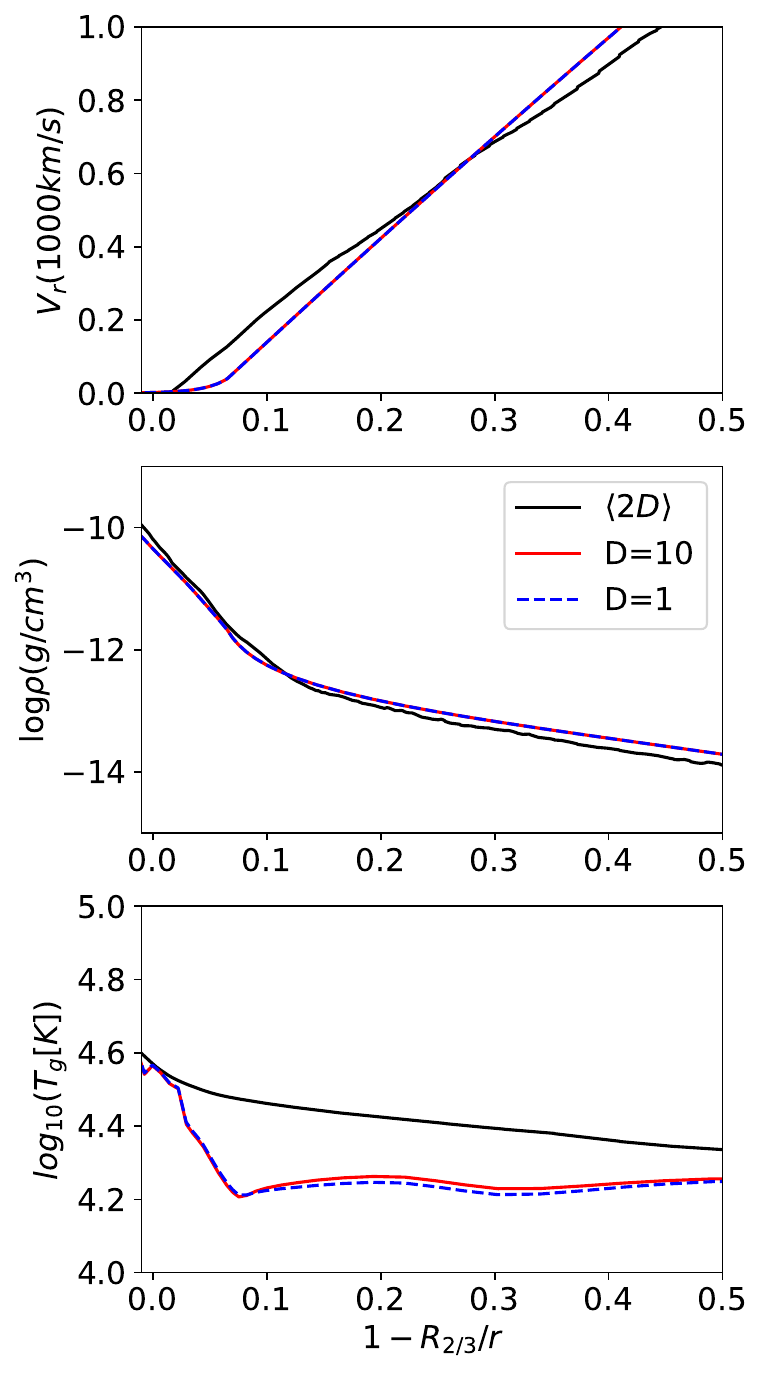}
      \caption{ Profile comparison for a smooth O4 model (dashed-blue) and a model with density contrast $D = 10$ (solid-red).  \textit{Upper panels: }Wind velocity profile. \textit{Middle panels: }Same as the \textit{upper panel} but for the density profile. \textit{Lower panel:} Same as the \textit{upper panel} but for the gas temperature. 
      }
      \label{fig:desconprofile}
  \end{figure}
      \begin{figure*}
      \centering
      \includegraphics[width=.75\linewidth]{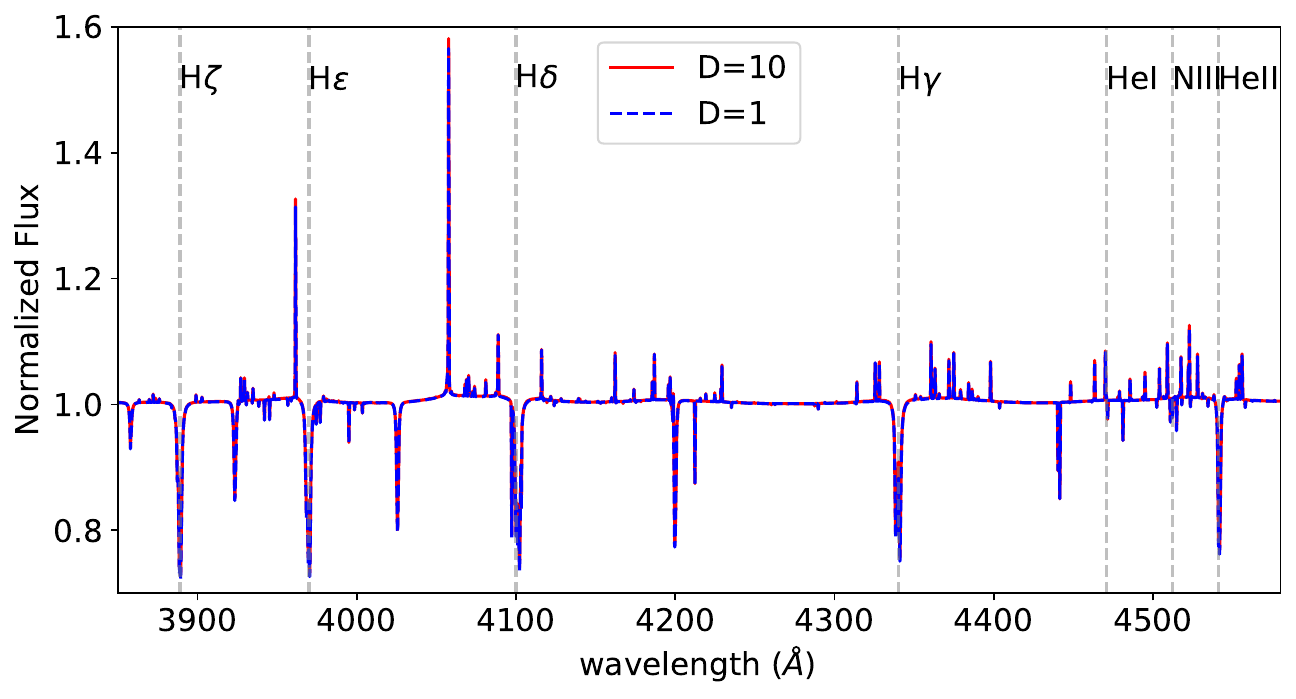}
      \caption{ Spectral line comparison for a smooth O4 model (dashed-blue) and a model with density contrast $D = 10$ (solid-red) in the optical. The rest of the parameters for both models are the same. }
      \label{fig:desconspectra}
  \end{figure*}
\section{Mass-weighted velocity profile}
 \begin{figure*}
      \centering
      \includegraphics[width=1.\linewidth]{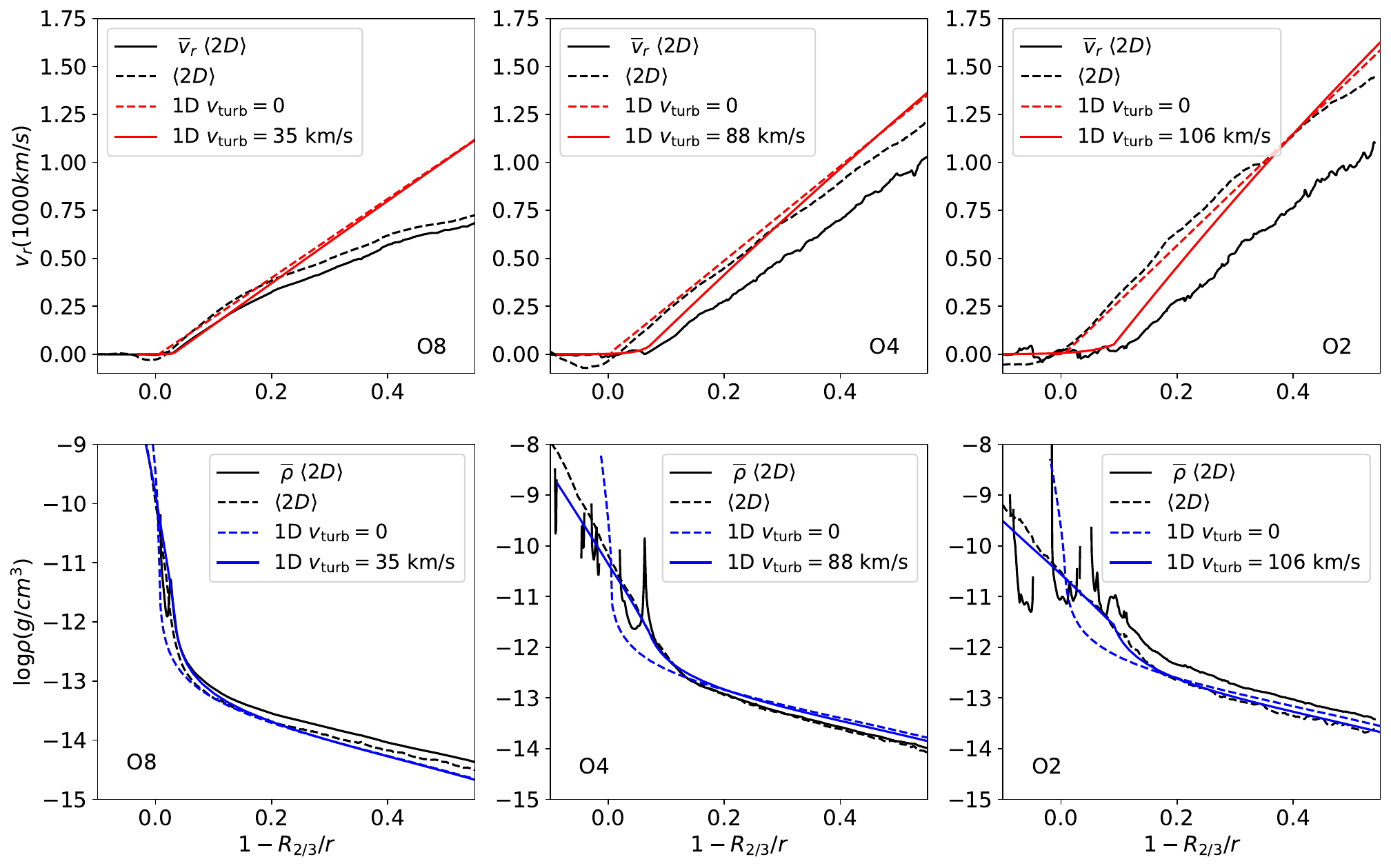}
      \caption{ Mass-weighted averaged 2D profiles. \textit{Upper panels:} Mass-weighted averaged 2D velocity ($\overline{v}_{r}$, solid black), spatial averaged 2D velocity profile (dashed black), the 1D PoWR model wind velocity profile with the best-fit parameters from Table~\ref{tab:params} and $\varv_{\mathrm{turb}}=0$ (dashed red) and for the same parameters but $\varv_{\mathrm{turb}}>0$ (solid red), from left to right for the O8, O4, O2 spectral types, respectively. \textit{Lower panels:} same as the upper panels but for the mass-weighted averaged 2D density profile. }
      \label{fig:massvel}
  \end{figure*}
 \section{Profile comparison with respect to $\tau_{\mathrm{Ross}}$}
   \begin{figure*}
      \centering
      \includegraphics[width=1.\linewidth]{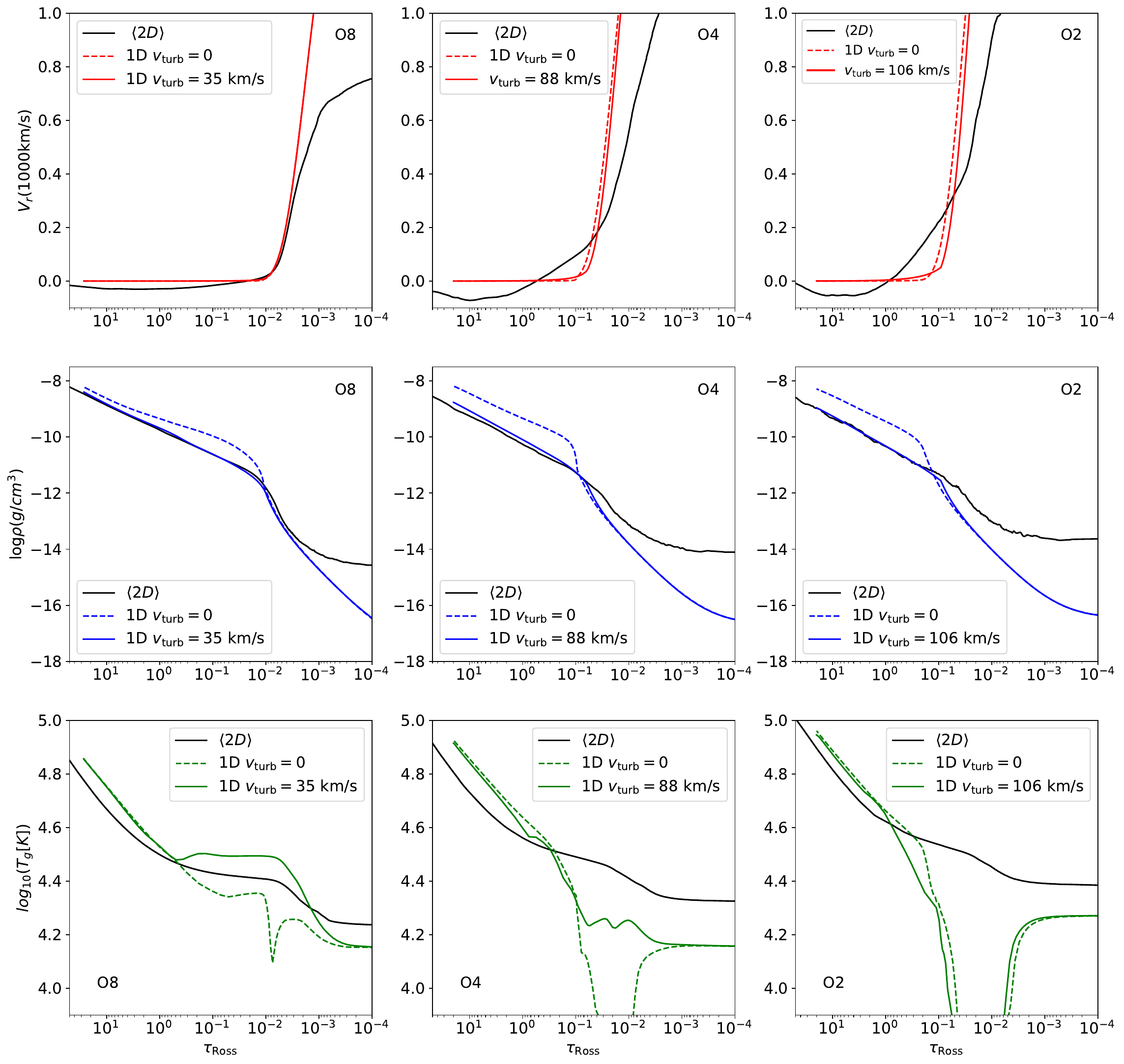}
      \caption{ Profile comparison for an O8, O4 and O2 stars with respect to the Rosseland mean opacity, $\tau_{\mathrm{Ross}}$. \textit{Upper panels: }Wind velocity profile, in solid black for the 2D averaged model of \citet{Debnath+2024}, dashed-red for the 1D PoWR model with the best-fit parameters from Table~\ref{tab:params} and $\varv_{\mathrm{turb}}=0$, and in solid red for the same parameters but $\varv_{\mathrm{turb}}>0$, from left to right for the O8, O4, O2 spectral types, respectively. \textit{Middle panels: }Same as the \textit{upper panel} but for the density profile. \textit{Lower panel:} Same as the \textit{upper panel} but for the gas temperature. }
      \label{fig:profile-taur}
  \end{figure*}
\section{Connection point, velocity gradient and radiative temperature}

        \begin{figure}
      \centering
      \includegraphics[width=.95\linewidth]{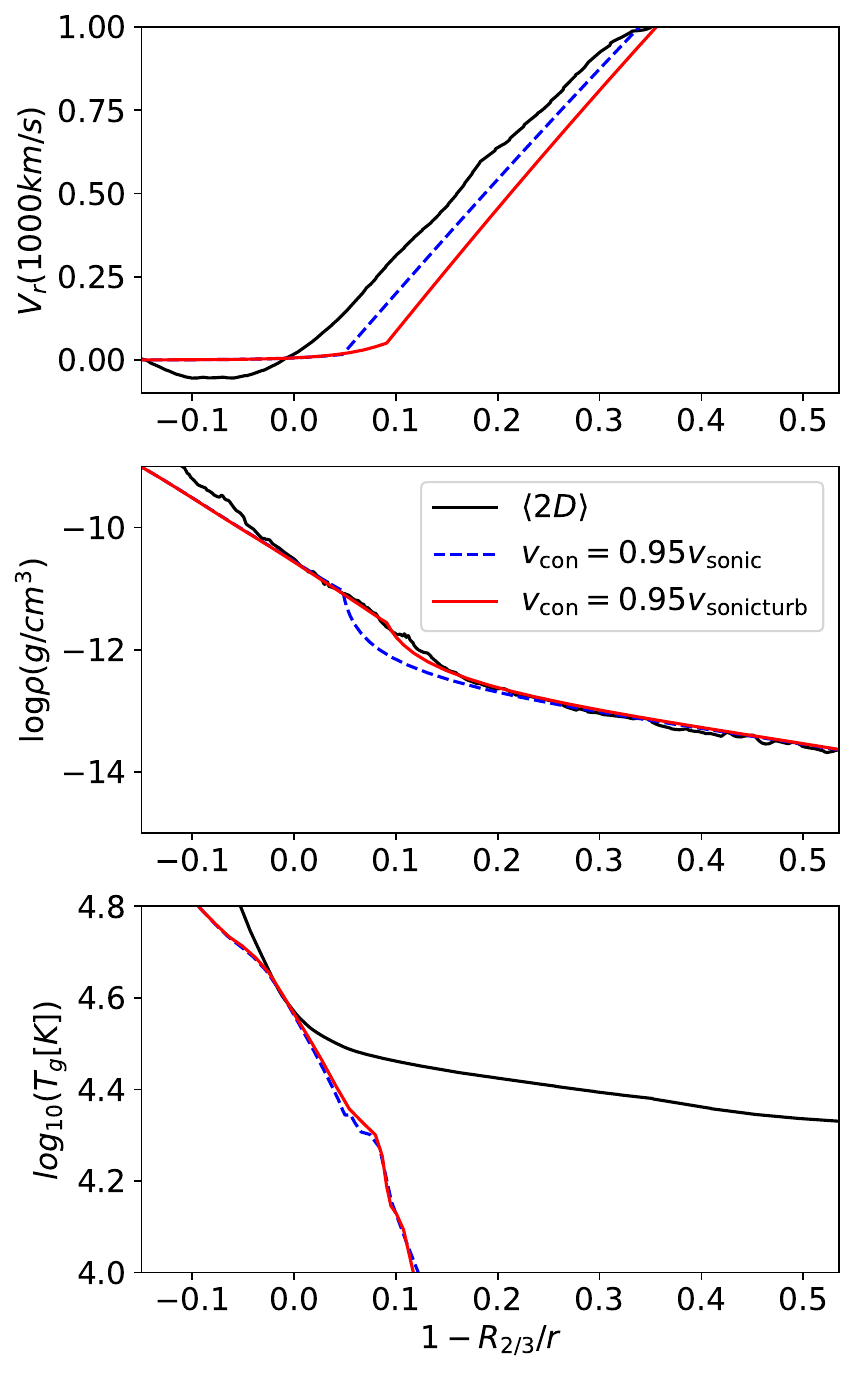}
      \caption{Profile comparison for the O2 model shifting the connection point between the (quasi-)hydrostatic regime and the $\beta-$law. \textit{Upper panel: }Wind velocity profile, in solid black for the 2D averaged model of \citet{Debnath+2024},  solid red for the best-fit 1D PoWR model with $\varv_{\mathrm{turb}}>0$ and initial connection point ($\varv_{\mathrm{con}}=0.95\,\varv_{\mathrm{sonicturb}}$), in dotted blue for the shifted connection point ($\varv_{\mathrm{con}}=0.95\,\varv_{\mathrm{sonic}}$). \textit{Middle panels: }Same as the \textit{upper panel} but for the density profile. \textit{Lower panel:} Same as the \textit{upper panel} but for the gas temperature.  }
      \label{fig:vsonic}
  \end{figure}
  
        \begin{figure}
      \centering
      \includegraphics[width=.95\linewidth]{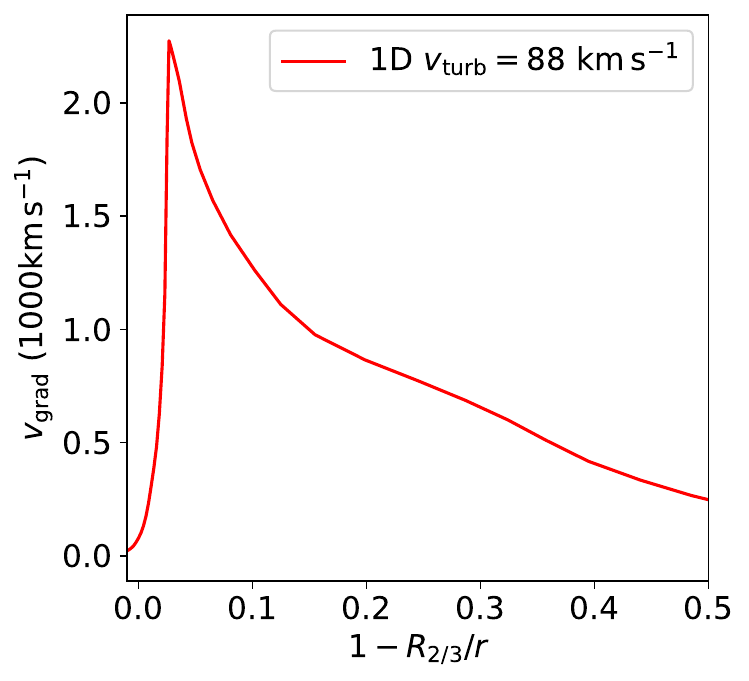}
      \caption{Velocity gradient for the O4 model with $\varv_{\mathrm{turb}}>0$. We can distinguish two different regimes: the hydrostatic regiome with $x\leq0.1$ and the $\beta-$law regime with $x>0.1$ }
      \label{fig:vgrad}
  \end{figure}
        \begin{figure}
      \centering
      \includegraphics[width=.95\linewidth]{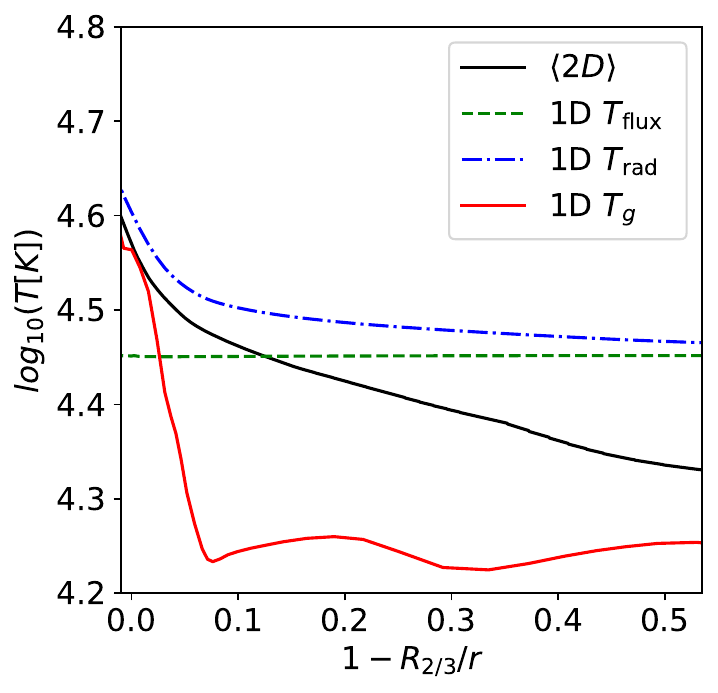}
      \caption{Flux temperature (dashed green line) and radiation temperature (dashed blue line) for the O4 model with $\varv_{\mathrm{turb}}>0$ obtained from the total emergent flux, compared to the gas temperature stratifications for the averaged $\langle$2D$\rangle$ and the 1D $\varv_{\mathrm{turb}}>0$ models. 
      }
      \label{fig:trad}
  \end{figure}

\section{$\xi$ broadening effect in the spectral synthesis}\label{app:xi}
        \begin{figure}
      \centering
      \includegraphics[width=1.\linewidth]{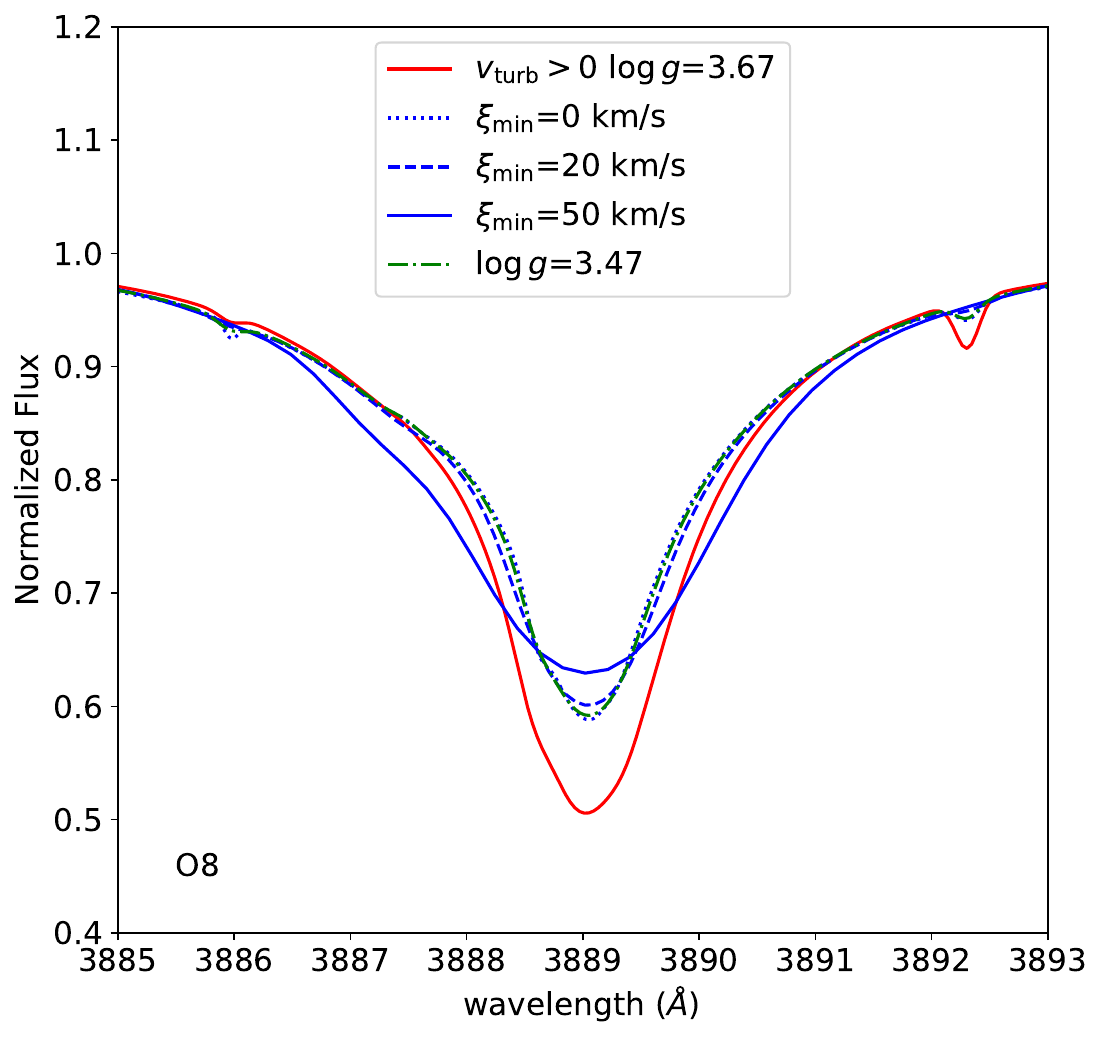}
      \caption{ Influence on the $\xi_{\mathrm{min}}$ microturbulence broadening term in the spectral synthesis for the H$\zeta$ line. In red the 1D fit with $\varv_{\mathrm{turb}}>0$, in dash-dotted green the model with $\varv_{\mathrm{turb}}=0$ and a lower $\log\,g$ with standard $\xi=10$ km\,s$^{-1}$, in blue the models with $\varv_{\mathrm{turb}}=0$ and a lower $\log\,g$ changing the $\xi_{\mathrm{min}}$ with 0, 20 and 50 km\,s$^{-1}$. The influence of $\xi_{\mathrm{min}}$ does not explain the effect on the $\varv_{\mathrm{turb}}>0$ for the depth of the lines. 
      }
      \label{fig:xi}
  \end{figure}
\section{Profile comparison for $\log g_{0}$}\label{app:log0}
      \begin{figure*}
      \centering
      \includegraphics[width=1.\linewidth]{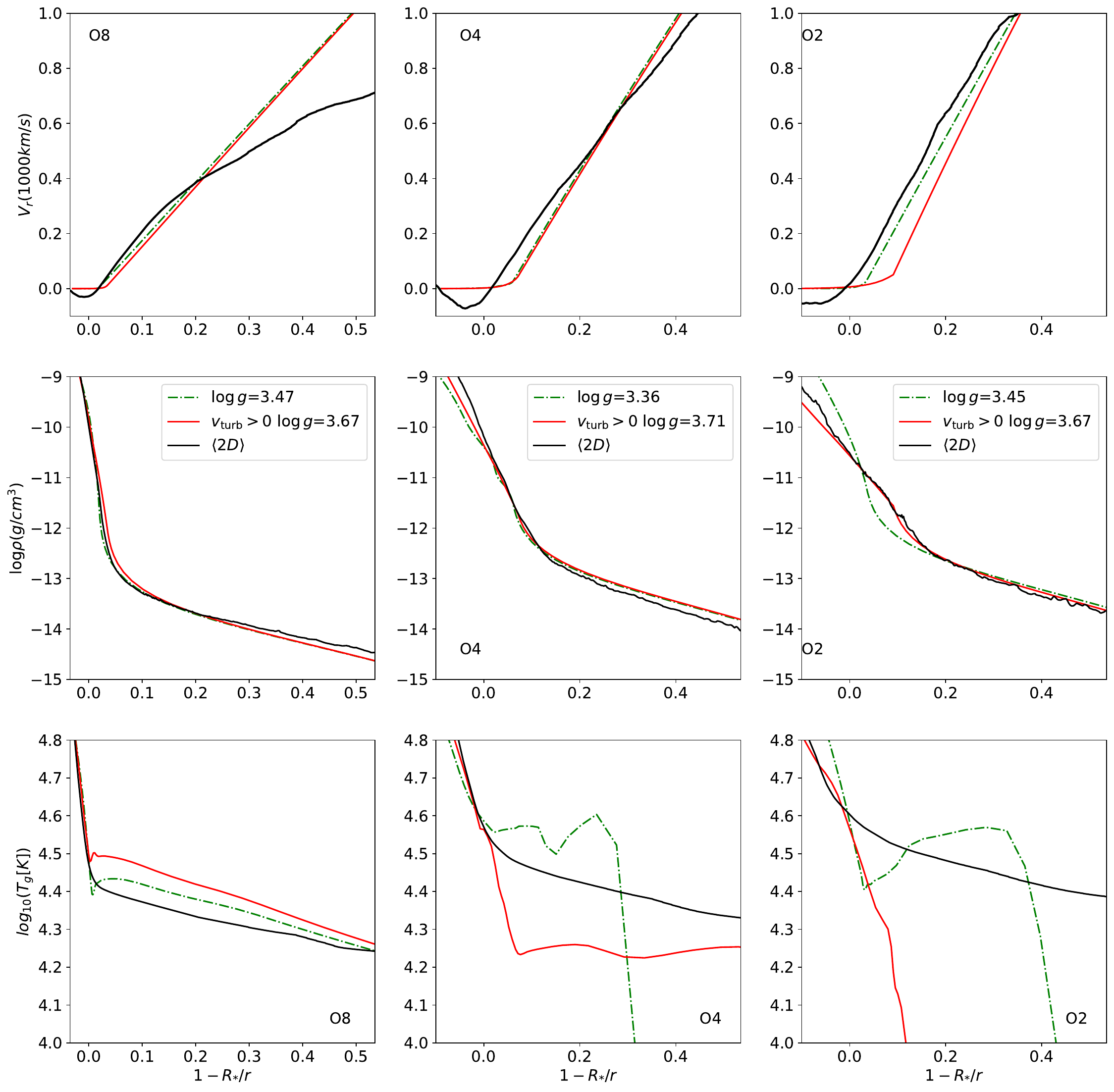}
      \caption{ Profile comparison for an O8, O4 and O2 stars. \textit{Upper panels: }Wind velocity profile, in solid black for the 2D averaged model of \citet{Debnath+2024},  solid red for the best-fit 1D PoWR model with $\varv_{\mathrm{turb}}>0$, in dashed green for $\varv_{\mathrm{turb}}=0$ and lower surface gravity, from left to right for the O8, O4, O2 spectral types, respectively. \textit{Middle panels: }Same as the \textit{upper panel} but for the density profile. \textit{Lower panel:} Same as the \textit{upper panel} but for the gas temperature.  }
      \label{fig:profile2}
  \end{figure*}

\section{Determination of evolutionary masses}\label{app:evolmasses}

\begin{figure*}
    \centering
    \includegraphics{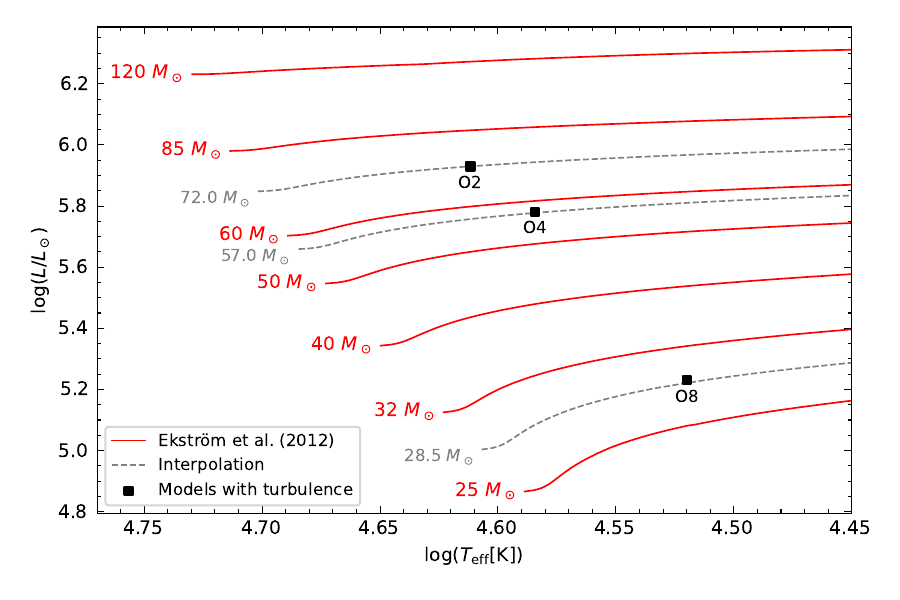}
    \caption{Hertzsprung-Russell diagram showing our models against the stellar evolution tracks of \citet{Ekstroem+2012} with the best-fit interpolated tracks.}
    \label{fig:hrd-evol}
\end{figure*}

To determine the evolutionary masses of our models, we plot their position in the HRD alongside the GENEC tracks at Galactic metallicity by \citet{Ekstroem+2012} in Fig\,\ref{fig:hrd-evol}. We perform a linear interpolation of the tracks in $\log (M_\mathrm{ini})$, and select the best fitting tracks by eye to the nearest $0.5M_\odot$ in initial mass. We then extract the current mass of the interpolated track at the point that lies closest to our data points in the HRD. The evolutionary masses obtained this way have an estimated error margin of $\pm 0.5M_\odot$ (without accounting for systematic errors in the evolution code) and are presented alongside the masses obtained in the atmosphere fitting in Table \ref{tab:mass}.

\section{Radiative acceleration}\label{app:acc}
\begin{figure}
      \centering
  \includegraphics[width=1.\linewidth]{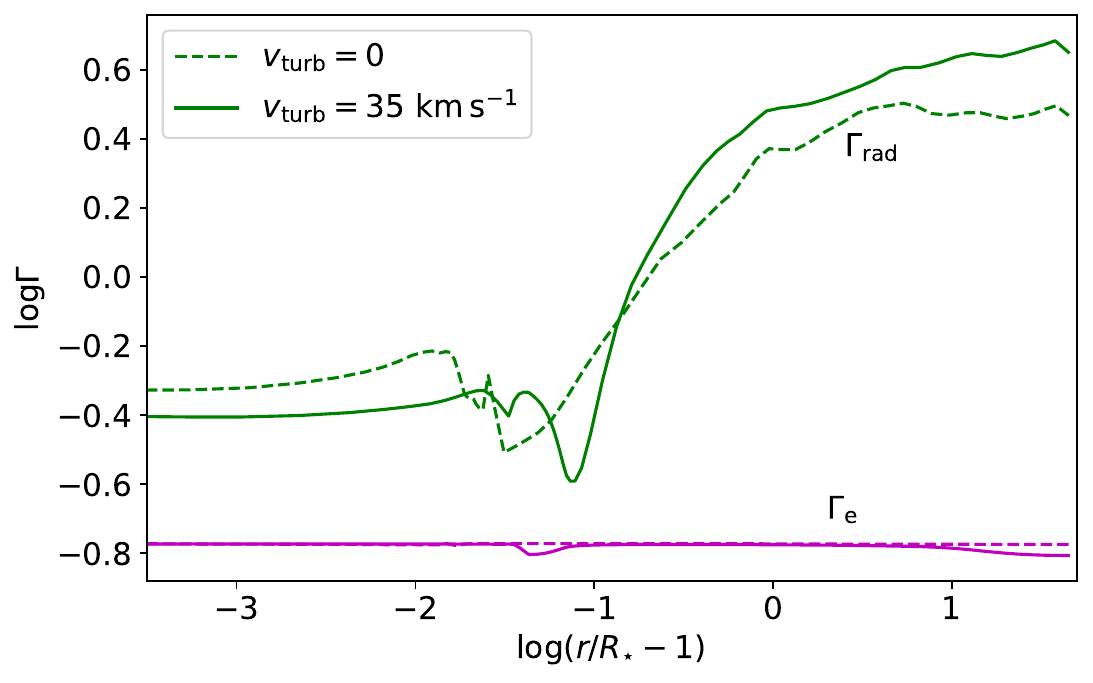}
  \caption{Same as Figure~\ref{fig:gamrad} but for the O8 model.}
  \label{fig:gamradO8}
\end{figure}
\begin{figure}
      \centering
  \includegraphics[width=1.\linewidth]{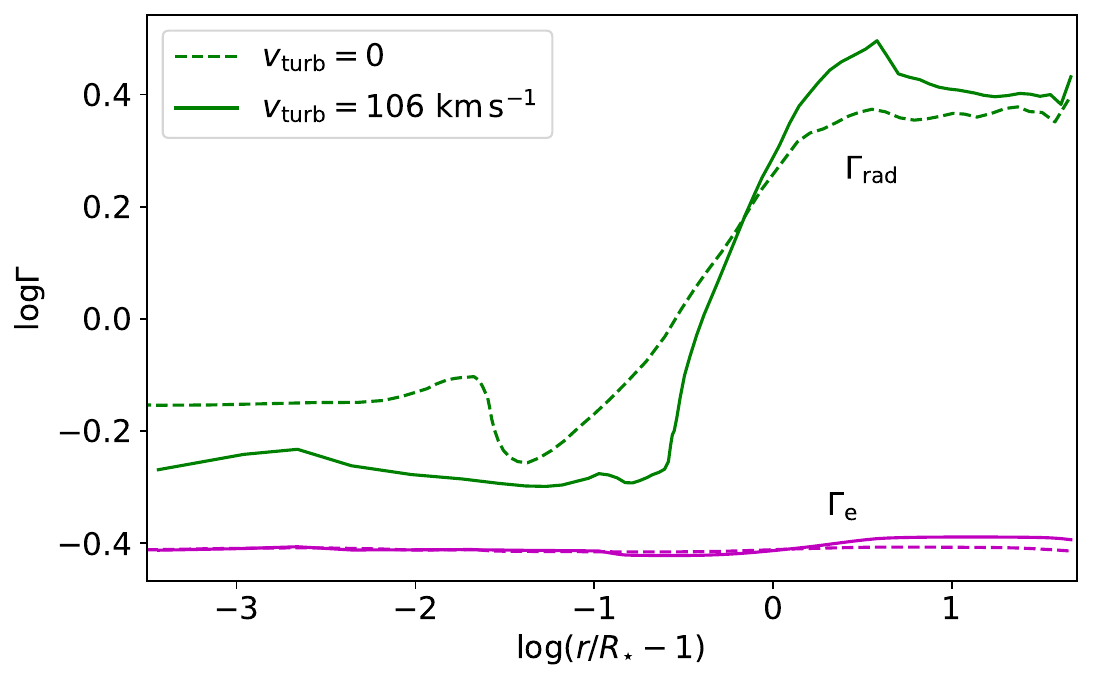}
  \caption{Same as Figure~\ref{fig:gamrad} but for the O2 model.}
  \label{fig:gamradO2}
\end{figure}
\end{appendix}
\end{document}